\documentclass[article, nojss]{jss}
\usepackage{amsmath}
\usepackage[utf8]{inputenc}
\usepackage{graphicx}
\usepackage{natbib}
\usepackage{enumerate}
\usepackage{microtype}
\usepackage{dsfont}
\usepackage{pifont}

\newcommand{\xmark}{\ding{55}}%
\newcommand{\cmark}{\ding{51}}%
\newcommand{\dnest}{\pkg{DNest4}}

\author{Brendon J. Brewer\\Department of Statistics\\The University of Auckland\And 
        Daniel Foreman-Mackey\\Sagan Fellow\\University of Washington}
\title{\pkg{DNest4}: Diffusive Nested Sampling in
\proglang{C++} and \proglang{Python}}

\Plainauthor{Brendon J. Brewer, Daniel Foreman-Mackey} %% comma-separated
\Plaintitle{DNest4: Diffusive Nested Sampling in
C++ and Python} %% without formatting
\Shorttitle{\pkg{DNest4}: Diffusive Nested Sampling} %% a short title (if necessary)

\Abstract{In probabilistic (Bayesian) inferences, we typically want to compute
properties of the posterior distribution, describing knowledge of
unknown quantities in the context of a particular dataset and the assumed
prior information. The marginal likelihood, also known as the ``evidence'',
is a key quantity in Bayesian model selection. The Diffusive Nested Sampling algorithm, a variant of Nested Sampling, is a powerful tool for generating
posterior samples and estimating marginal likelihoods. It is effective at
solving complex problems including many where the posterior distribution is
multimodal or has strong dependencies between variables.
\pkg{DNest4} is an open source (MIT licensed),
multi-threaded implementation of this algorithm in
\proglang{C++11},
along with associated utilities including:
i)
\pkg{RJObject}, a class template for finite mixture models;
(ii)
A \proglang{Python} package allowing basic use without \proglang{C++} coding; and
iii)
Experimental support for models implemented in \proglang{Julia}.
In this paper we demonstrate \pkg{DNest4} usage through examples including
simple Bayesian data analysis, finite mixture models, and Approximate
Bayesian Computation.
}

\Plainkeywords{bayesian inference, markov chain monte carlo,
metropolis algorithm, bayesian computation, nested sampling, c++11, python} %% without formatting
\Address{
  Brendon J. Brewer\\
  Department of Statistics\\
  The University of Auckland\\
  Private Bag 92019\\
  Auckland, 1142\\
  New Zealand\\
  E-mail: \email{bj.brewer@auckland.ac.nz}\\
  URL: \url{https://www.stat.auckland.ac.nz/~brewer/}
}

\newcommand{\params}{\theta}
\newcommand{\data}{D}

\begin{document}
\maketitle

%% Need this after the abstract
%\setlength{\parindent}{0pt}
%\setlength{\parskip}{8pt}

\section{Introduction}
Bayesian inference, where probability theory describes degrees of
logical implication or subjective certainty, provides a powerful general basis
for data analysis \citep{o2004kendall, sivia2006data}. The result of such
an analysis is typically
posterior probabilities of various hypotheses, or
a joint posterior probability distribution for the values of unknown
parameters.

In compact but standard notation, the posterior
posterior distribution for parameters $\params$ given data $\data$, within
the context of prior information $M$, is
\begin{align}
p(\params | \data, M) &=
\frac{p(\params | M)p(\data | \params, M)}{p(\data | M)}
\end{align}
or
\begin{align}
\textnormal{posterior} &=
\frac{\textnormal{prior} \times \textnormal{likelihood}}
     {\textnormal{marginal likelihood}}.
\end{align}

If prior information $I$ (dropped hereafter)
implies a set of possible `models' $\{M_i\}$,
rather than a single one $M$, the posterior model probabilities are given by
\begin{align}
P(M_i | \data) &=
\frac{P(M_i)p(\data | M_i)}{\sum_j P(M_j)p(\data | M_j)}
\end{align}
where
\begin{align}
p(\data | M_j) &= \int p(\theta_j | M_j)p(\data | \theta_j, M_j) \, d\theta_j
\end{align}
is the marginal likelihood of model $j$, equal to the expected value of the
likelihood function with respect to the prior distribution.
This kind of calculation is often
called ``model selection'' or ``model averaging'', and the results
are often presented as ratios
of marginal likelihoods, known as Bayes Factors.
When discussing computational matters, the prior distribution for parameters
is
often written $\pi(\theta)$, the likelihood $L(\theta)$,
and the marginal likelihood $Z$. A popular alternative name for the marginal
likelihood, which emphasizes its role in Bayesian model averaging,
is the ``evidence''.

Nested Sampling \citep[NS;][]{skilling2006nested} is a Monte Carlo method whose main
aim is to calculate $Z$. However, it can also be used to generate samples
to represent the posterior distribution $\pi(\theta)L(\theta)/Z$, or
any other distribution proportional to
$\pi(\theta)\Phi\left[L(\theta)\right]$ where $\Phi$ is any monotonic function.
This latter property makes NS particularly useful for statistical mechanics
calculations \citep{partay2010efficient, baldock2016determining},
where the
``canonical'' family of distributions proportional to
$\pi(\theta)L(\theta)^\beta$ is of interest. In such applications,
$L(\theta)$ is usually equivalent to
$\exp(-\textnormal{energy})$. NS is particularly efficient for this, since
only a single run (exploring regions of higher and higher $L$) is needed, and different canonical
distributions can be obtained by re-weighting the output points.

A defining feature of NS is that it works with a sequence of
{\em constrained prior} distributions, proportional to $\pi$ but
restricted to regions of the parameter space where $L(\theta)$
is above some threshold $\ell$:
\begin{align}
p(\theta; \ell) &=
\frac{\pi(\theta)\mathds{1}\left[L(\theta) > \ell\right]}{X(\ell)}
\label{eqn:constrained_prior}
\end{align}
where
\begin{align}
X(\ell) &= \int \pi(\theta) \mathds{1}\left[L(\theta) > \ell\right] \, d\theta
\end{align}
is the amount of prior mass which has likelihood greater than $\ell$, and
$\mathds{1}()$ is the indicator function which takes the value 1 if the
argument is true and 0 otherwise.
In the standard NS framework, the sequence of $\ell$ values is selected
so that $X(\ell)$ shrinks by a factor $\approx e^{-1/N}$ per iteration, where
$N$ is the number of particles used. This geometric compression of the
parameter space is a defining feature of NS.

Sampling from the constrained priors (Equation~\ref{eqn:constrained_prior})
is required, and Markov Chain Monte Carlo (MCMC) is a popular method of doing
this, although alternatives exist
\citep[e.g.,][]{feroz2009multinest, handley2015polychord}.
Diffusive Nested Sampling \citep[DNS;][]{brewer2011diffusive}
is an alternative to NS for
problems where MCMC is the only viable sampling method. DNS is based on the
Metropolis algorithm, and evolves one or more particles in the parameter space, along with
an integer index variable $j$ for each particle,
to explore the following joint distribution:
\begin{align}
p(\theta, j) &= p(j)p(\theta | j)\\
&= w_j \times
\frac{\pi(\theta)\mathds{1}\left[L(\theta) > \ell_j\right]}{X(\ell_j)}.
\label{eqn:target_distribution}
\end{align}
where the $\{\ell_j\}$ are a sequence of increasing likelihood thresholds
or {\em levels},
$\ell_0 = 0$, and $\{w_j\}$ is the marginal distribution for $j$.
The marginal distribution for $\theta$ is then a {\em mixture} of
constrained priors:
\begin{align}
p(\theta) &=
\pi(\theta)\sum_{j=0}^{j_{\rm max}}
\frac{w_j\mathds{1}\left[L(\theta) > \ell_j\right]}{X(\ell_j)}.
\label{eqn:mixture_of_constrained_priors}
\end{align}
The DNS algorithm consists of two stages. In the first stage,
the particle(s) are initialized from the prior $\pi(\theta)$, equivalent
to the mixture of constrained priors with a single level whose log likelihood
threshold is $-\infty$. The mixture of constrained priors evolves by adding new levels,
each of which compresses the distribution by a factor of about $e \approx 2.71818$
(see \citet{brewer2011diffusive} for details). In this stage,
the mixture weights $\{w_j\}$ are set according to
\begin{align}
w_j &\propto \exp(j/\lambda),\label{eqn:weighting}
\end{align}
where $\lambda$ is a scale length.
This enhances the probability that particles stay close to the
highest likelihood regions seen.
The second stage sets the mixture weights to be approximately
uniform ($w_j \propto 1$), with some tweaks described by
\citet{brewer2011diffusive}.

The mixture of constrained priors tends to be easier to sample than the
posterior, as the prior is always a mixture component, allowing the
MCMC chain to mix between different modes in some circumstances. The marginal
likelihood estimate can also be more accurate than standard MCMC-based NS
\citep{brewer2011diffusive}, as less information is discarded.
DNS has been applied several times in astrophysics
\citep[e.g.,][]{pancoast2014modelling, huppenkothen2015dissecting,
brewer2015fast}
and was recently used in a biological application
\citep{dybowski2015single}.

There are several ways of using \pkg{DNest4}. After installing the software
(Section~\ref{sec:installation}), you can implement a model as
a \proglang{C++} class
(Section~\ref{sec:models}) and compile it to create an executable file
to run \pkg{DNest4} on that problem. This method offers full control over the
design of your class, and allows the opportunity of optimizing
performance by preventing a full re-computation of the log-likelihood when
only a subset of the parameters has been changed.

Alternatively, the \proglang{Python}
bindings (Section~\ref{sec:python_bindings}) allow you
to specify a model class in \proglang{Python}, and run \pkg{DNest4} entirely
in the \proglang{Python} interpreter without having to invoke the \proglang{C++}
compiler.

\section{Relation to other algorithms}\label{sec:relation}
The Diffusive Nested Sampling algorithm, and its implementation in
\pkg{DNest4}, has advantages and disadvantages compared to other Bayesian
computation algorithms and software.

In the Nested Sampling sphere,
the simple \proglang{C} implementation of given by
\citet{skilling2006nested} is useful for understanding the classic NS
algorithm. Similarly to \pkg{DNest4}, it is the user's responsibility
to implement the MCMC moves used. The DNS algorithm produces more
accurate estimates of $Z$ and ought to outperform classic NS on multimodal
problems.

\pkg{MultiNest} \citep{feroz2009multinest} is a
\proglang{Fortran}
Nested Sampling implementation
that dispenses with MCMC for generating new particles from constrained
prior distributions. Instead, the new particles are generated by making
an approximation to the constrained prior using ellipsoids. This tends to work
well in low to moderate dimensional parameter spaces (up to $\sim 40$).
In higher dimensions, \pkg{DNest4} is more useful than \pkg{MultiNest}
since it is based on MCMC. We expect \pkg{MultiNest} to be more useful
than
\pkg{DNest4} only on low to moderate dimensional problems with slow
likelihood evaluations, where \pkg{DNest4}'s reliance on MCMC becomes costly.
The more recent \pkg{POLYCHORD} \citep{handley2015polychord} combines
\pkg{MultiNest}-like methods with MCMC.

Outside of Nested Sampling,
the popular \pkg{JAGS} \citep{jags} and \pkg{Stan} \citep{stan} packages
are more convenient than \pkg{DNest4} for specifying models, by virtue
of their model-specification languages. For users who are interested only
in the posterior distribution and do not need the marginal likelihood,
these are very useful.
However, being based on Gibbs
sampling and Hamiltonian MCMC respectively, they can run into difficulty
on multimodal posterior distributions.

\pkg{emcee} \citep{emcee} is a popular \proglang{Python} package for MCMC
based on affine-invariant ensemble sampling. The user needs only to
specify a \proglang{Python} function evaluating the log of the posterior
density (up to a normalizing constant). This user-friendliness is a key
advantage of \pkg{emcee}, and its algorithm performs well on low--moderate
dimensional (up to $\sim 50$) parameter spaces with a unimodel
target distribution (which can be highly dependent). However, it
can give misleading results in higher dimensions \citep{huijser}.

Algorithms based on ``annealing'' or ``tempering'', such as parallel
tempering \citep{hansmann1997parallel} and annealed importance sampling
\citep{neal2001annealed} are
related to Nested Sampling and are useful on multimodal posterior distributions.
However, they require much more tuning than NS (in the form of an
{\em annealing schedule}) and do
not work on phase change problems, unlike Nested Sampling
\citep{skilling2006nested}.

The advantages and disadvantages of a subset of these
packages are summarized in Table~\ref{tab:packages}.

%\begin{table}
%
%  \begin{tabular}{cc}
%    Knuth & Lamport
%  \end{tabular}}
%\end{table}

\begin{table}
\begin{center}
\resizebox{\textwidth}{!}{%
\begin{tabular}{|l|c|c|c|c|c|c|}
\hline
Package     &   Easy to     & High   &   Multimodal     & Dependent   &   Phase   &   Computes $Z$?\\
            &   implement & dimensions? & distributions? & distributions? & changes? & \\
& models? & & & & &\\
\hline
\pkg{DNest4}    & \xmark  & \cmark & \cmark & \cmark &\cmark &\cmark\\
\pkg{emcee}    & \cmark  & \xmark & \xmark & \cmark &\xmark &\xmark\\
\pkg{JAGS}    & \cmark  & \cmark & \xmark & \xmark & \xmark &\xmark\\
\pkg{MultiNest}    & \cmark  & \xmark & \cmark & \cmark & \cmark &\cmark\\
\pkg{Stan}    & \cmark  & \cmark & \xmark & \cmark & \xmark &\xmark\\
\hline
\end{tabular}}
\caption{A simplified summary of the advantages and disadvantages of some
Bayesian computation software packages.\label{tab:packages}}
\end{center}
\end{table}

\section{Markov chain monte carlo}\label{sec:mcmc}
DNS is build upon the Metropolis-Hastings algorithm.
In this algorithm, the acceptance probability $\alpha$
is given by
\begin{align}
\alpha &= \min\left(1,
\frac{q(\params'|\params)}{q(\params | \params')}
\frac{\pi(\params')}{\pi(\params)}\frac{L(\params')}{L(\params)}
\right)
\end{align}
where $q(\theta' | \theta)$ is the proposal distribution used to generate
a new position $\theta'$ from the current position $\theta$. Often,
$q$ is {\em symmetric} so the $q$ terms cancel in the acceptance probability.

In DNS, the target distribution is not the posterior but rather
the joint distribution in Equation~\ref{eqn:target_distribution}.
Moves of $\theta$ are done keeping $j$ fixed, so we only need
to consider the Metropolis acceptance probability for fixed $j$,
i.e., with respect to a single constrained prior like
Equation~\ref{eqn:constrained_prior}.
Hence, the appropriate acceptance probability
for a proposed move from $\theta$ to $\theta'$ is
\begin{align}
\alpha &= \min\left[1,
\frac{q(\params'|\params)}{q(\params | \params')}
\frac{\pi(\params')}{\pi(\params)}
\mathds{1}\left(L(\params') > \ell_j\right)
\right]
\label{eqn:log_hastings}
\end{align}
where $\ell_j$ is the likelihood threshold for the current level $j$.
There are also moves that propose a change to $j$ while keeping
$\params$ fixed, but the details are less relevant to the user.

For convenience later on, we 
separate the prior and proposal-related terms from
the likelihood-related term, and write the former as
\begin{align}
H = \frac{q(\params'|\params)}{q(\params | \params')}
\times \frac{\pi(\params')}{\pi(\params)}.
\end{align}

The logarithm of $H$,
\begin{align}
\ln(H) = \ln\left[\frac{q(\params'|\params)}{q(\params | \params')}
\times \frac{\pi(\params')}{\pi(\params)}\right],
\end{align}
is the user's responsibility when implementing models,
and will become relevant in Sections~\ref{sec:perturb}
and~\ref{sec:proposals}.
In terms of $\ln(H)$, the
acceptance probability becomes
\begin{align}
\alpha &= \min\left[1,
e^{\ln(H)}\times
\mathds{1}\left(L(\params') > \ell_j\right)
\right].\label{eqn:logH}
\end{align}

%The \pkg{DNest4} \code{Sampler} class handles the likelihood check,
%while the \code{perturb(DNest4::RNG\&)} member function handles the
%rest of the terms. The value returned from \code{perturb(DNest4::RNG\&)}
%must be
%for your problem.

%\subsection{Diffusive Nested Sampling}

\section{Dependencies and installation}\label{sec:installation}
The following instructions apply to Unix-like operating systems such as
GNU/Linux, Mac OS X, and FreeBSD. Currently we have not tested
\pkg{DNest4} on Microsoft Windows.

Development of \pkg{DNest4} takes place in the \pkg{git} repository located at
\begin{center}
\url{https://github.com/eggplantbren/DNest4/}
\end{center}
The software
is licensed under the permissive open source
MIT license. To compile and run \pkg{DNest4},
you require a recent version of the GNU
\proglang{C++} compiler, \pkg{g++} \citep{gcc}.
\pkg{DNest4} uses features from the \proglang{C++11} standard
\citep{c++11}.

The \proglang{Python}
packages \pkg{NumPy} \citep{numpy}, \pkg{matplotlib} \citep{matplotlib},
and \pkg{Cython} \citep{cython} are also needed.
To download and compile \pkg{DNest4}
the following steps are sufficient:
\begin{CodeChunk}
\begin{CodeInput}
> wget https://github.com/eggplantbren/DNest4/archive/0.1.4.tar.gz
> tar xvzf 0.1.4.tar.gz
> mv DNest4-0.1.4 DNest4
> cd DNest4/code
> make
> cd ../python
> python setup.py install
\end{CodeInput}
\end{CodeChunk}

In Mac OS X, the final line (which installs the \proglang{Python} parts)
of \pkg{DNest4})
needs to provide information about
your OS version. For example, if your computer runs Mac OS X 10.9,
the installation command for the \proglang{Python} package is
\begin{CodeChunk}
\begin{CodeInput}
> MACOSX_DEPLOYMENT_TARGET=10.9 python setup.py install
\end{CodeInput}
\end{CodeChunk}

We recommend you create an environment variable called \code{DNEST4\_PATH}
and set it to the directory {\em above} the \pkg{DNest4} directory. Then,
if the model templates from the \pkg{DNest4} repository are copied
to any other location on your system and used as the basis for
new work, their Makefiles will continue to function.

\section[Running DNest4]{Running \pkg{DNest4}}\label{sec:running}
To demonstrate \pkg{DNest4},
we will use a simple linear regression example where the
sampling distribution is
\begin{align}
y_i | m, b, \sigma &\sim \textnormal{Normal}(mx_i + b, \sigma^2)
\end{align}
and the priors are
\begin{align}
m &\sim \textnormal{Normal}(0, 1000^2)\\
b &\sim \textnormal{Normal}(0, 1000^2)\\
\ln\sigma &\sim \textnormal{Uniform}(-10, 10).
\end{align}
These are naive diffuse priors and do not have any special status.
The dataset is shown in Figure~\ref{fig:regression_lines}
and the code is included in the {\tt code/Examples/StraightLine}
subdirectory, and a slightly simplified version of the
code is explained in Section~\ref{sec:models}.
To execute \pkg{DNest4} on this problem, go to this directory and
execute {\tt main}. The output to the screen
should contain information about levels and ``saving particles to disk''.
After 10,000 particles have been saved to the disk, the run will terminate.

\section{Output files}
The executable {\tt main} is responsible for the exploration part of the
DNS algorithm (i.e., running the MCMC chain, building
levels, and then exploring all the levels). It creates three text output files,
{\tt sample.txt}, {\tt sample\_info.txt}, and {\tt levels.txt}.

The first output
file, {\tt sample.txt}, contains a sampling of parameter values that
represents the {\it mixture of constrained priors} (the target distribution
used in DNS), {\bf not} the
posterior distribution. Each line of {\tt sample.txt} represents a point in
parameter space. In the linear example, there are three parameters
($m$, $b$, and $\sigma$), so there
are three columns in {\tt sample.txt}.
Each time a point is saved to {\tt sample.txt}, \pkg{DNest4} prints
the message ``Saving a particle to disk. N = ...''.

The second output file, {\tt sample\_info.txt}, should have the same number of
rows as {\tt sample.txt}, because it contains metadata about the samples in
{\tt sample.txt}. The first
column is the index $j$, which tells us which ``level'' the particle was in
when it was saved. Level 0 represents the prior, and higher levels represent
more constrained versions of the prior.
The second column is the log-likelihood value, and the third column is
the likelihood ``tiebreaker'', which allows Nested Sampling to work when
there is a region in parameter space with nonzero prior probability where the
likelihood is constant. The final column tells us which thread the particle
belonged to: when you use \dnest~in multithreaded mode
(see Appendix~\ref{sec:command_line_options}), each thread
is responsible for evolving one or more particles.

The third output file, {\tt levels.txt}, contains information about the levels
that were built during the run. The first column has estimates of the $\log(X)$
values of the levels, i.e., how compressed they are relative to the prior, in
units of nats. For example, if a level has $\log(X) = -1.02$, its likelihood
value encloses $\exp(-1.02) \approx 36.1\%$ of the prior mass.

The second column contains the log likelihoods of the levels.
The first level, with a $\log(X)$ value of 0 and a log likelihood of
$-10^{308}$ (basically ``minus infinity''), is simply the prior. The third
column has the ``tiebreaker'' values for the levels, which again are not
particularly useful unless your problem has likelihood plateaus. The fourth
and fifth columns are the number of accepted proposals and the total number
of proposals that have occurred within each level, which are useful for
monitoring the Metropolis acceptance ratio as a function of level.
The final two columns, called ``exceeds'', and ``visits'', are used to refine
the estimates of the level compressions (and hence the $\log(X)$ values of
the levels in column 1), as discussed in Section 3 of
\citet{brewer2011diffusive}.
The visits column counts the number of times a level (level $j$, say)
has been visited, but only starts counting after the next level ($j+1$) has been created. The exceeds column counts the number of times a particle that was
in level $j$ had a likelihood that exceeded that of level $j+1$.

\section{Postprocessing}\label{sec:postprocessing}
\begin{figure}[ht!]
\begin{minipage}{7.5cm}
\centering
\includegraphics[width=8cm]{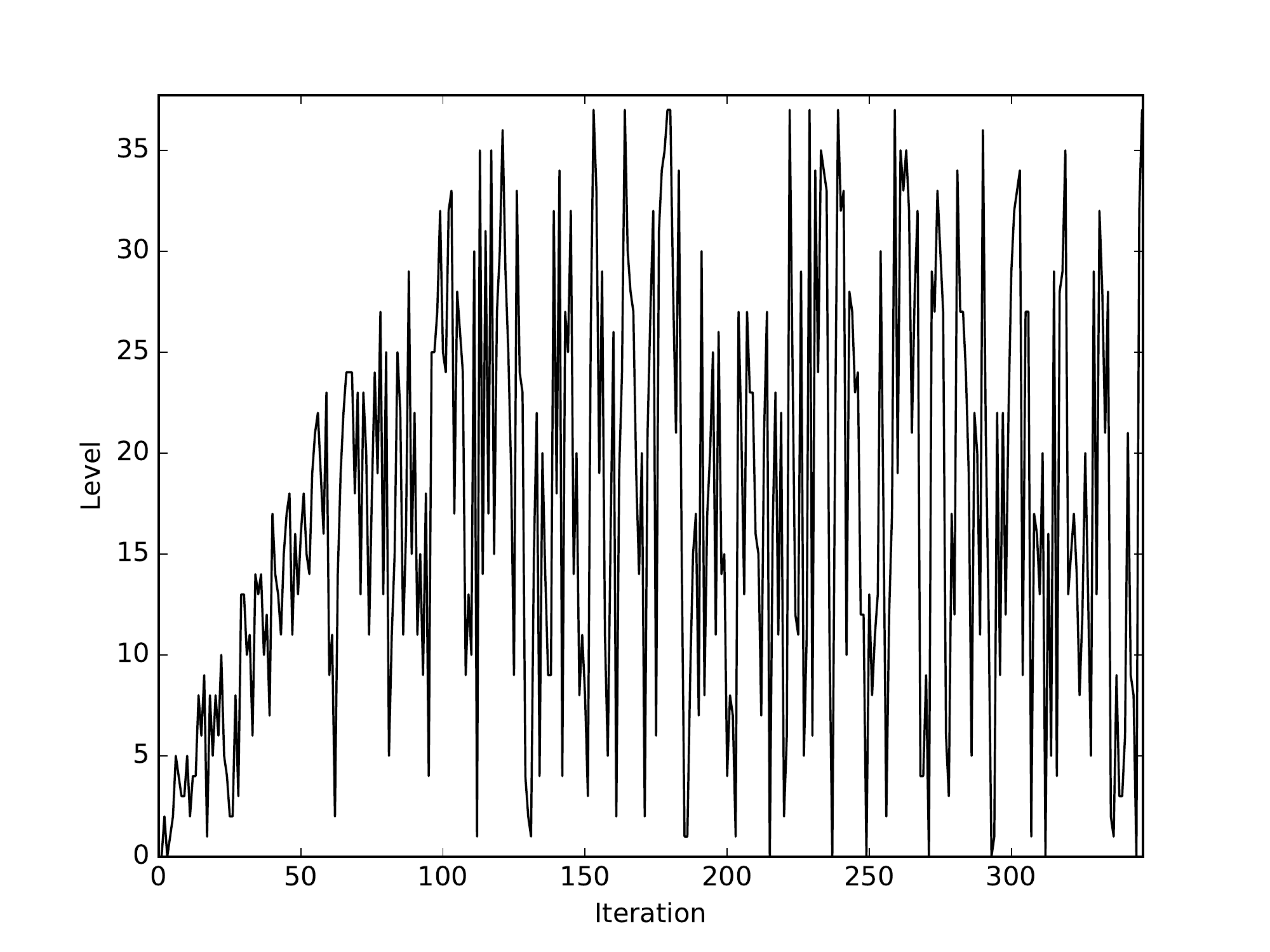}
\caption{The level $j$ of the saved particles over time.
Typically, this will trend upwards until all the levels have
been created, and then diffuse evenly throughout all the levels.
\label{fig:fig1}}
\end{minipage}\hspace{0.5cm}
\begin{minipage}{7.5cm}
\centering
\includegraphics[width=8cm]{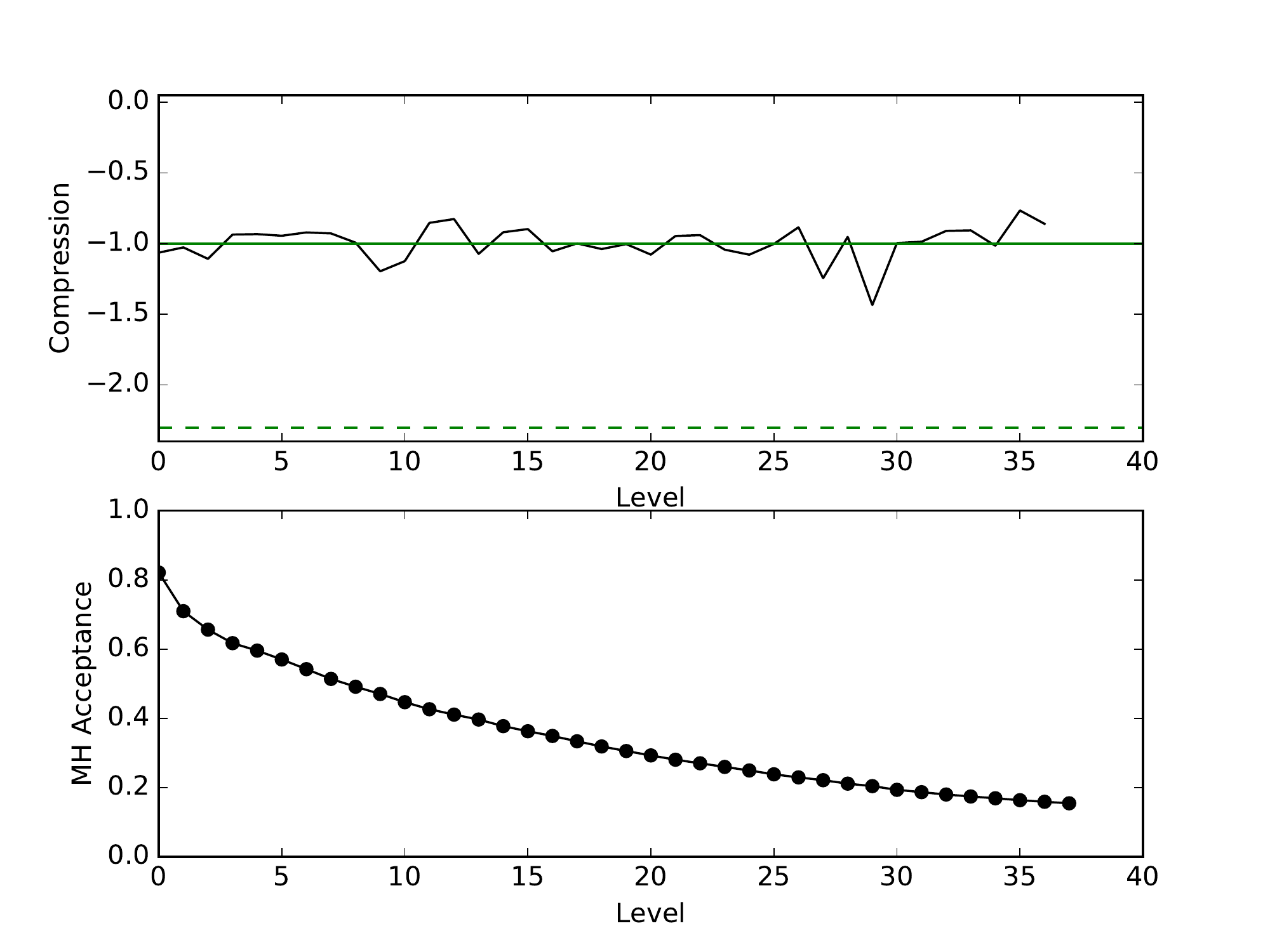}
\caption{Top panel: The estimated compression factor between subsequent
levels, expressed as $\ln(X_{i+1}/X_{i})$. Bottom panel:
The Metropolis acceptance fraction as a function of level.
\label{fig:fig2}}
\end{minipage}
\end{figure}

\begin{figure}[ht!]
\begin{minipage}{7.5cm}
\includegraphics[width=7cm]{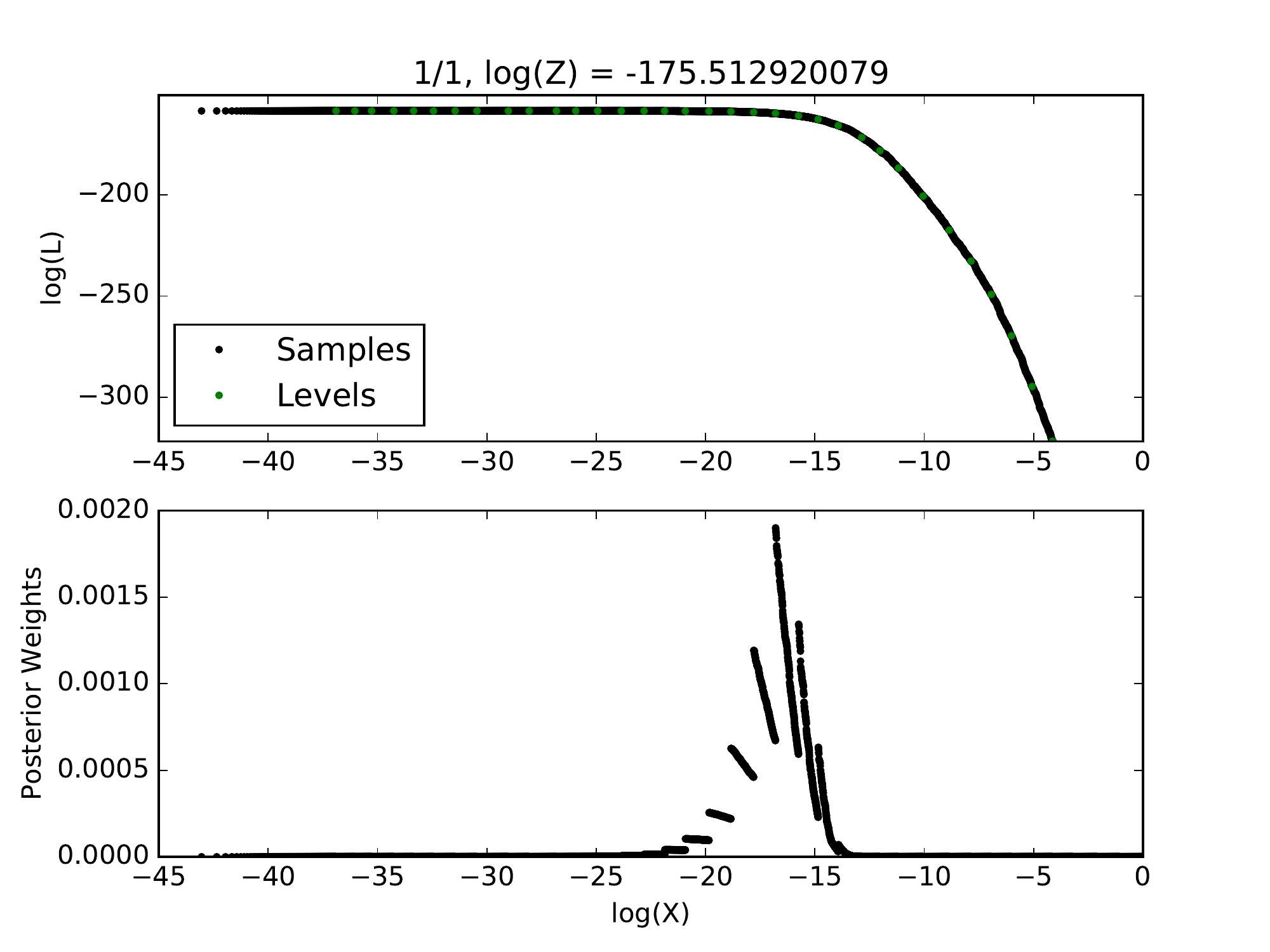}
\caption{Top panel: The log-likelihood curve, showing the relationship
between log-likelihood and the enclosed prior mass.
Bottom panel: Posterior weights of the saved particles.
For a successful run, there should be a clear peak, and saved particles
to the left of this plot should have insignificant posterior weight compared
to those in the peak.
\label{fig:fig3}}
\end{minipage}\hspace{0.5cm}
\begin{minipage}{7.5cm}
\includegraphics[width=7cm]{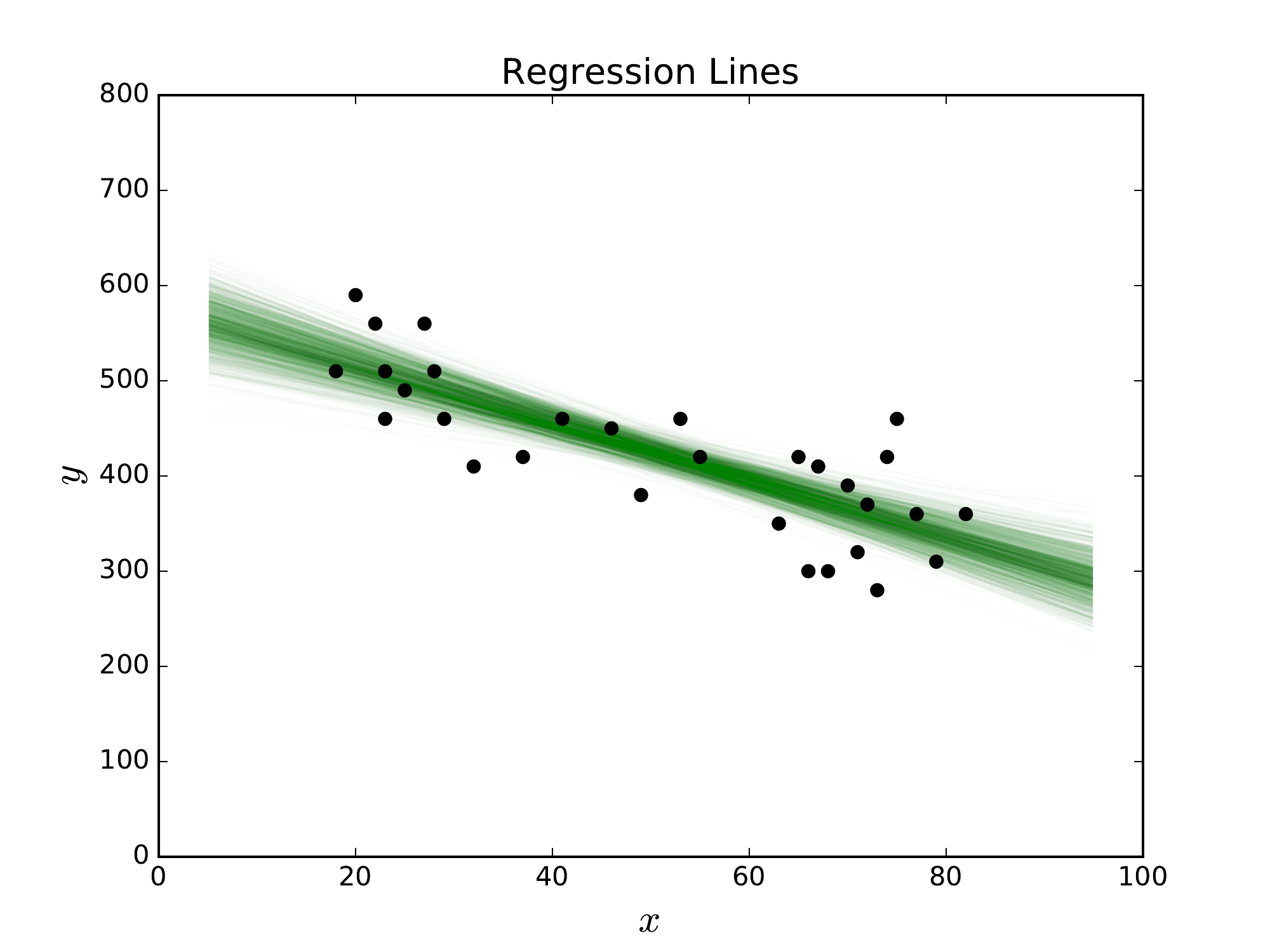}
\caption{Regression lines drawn from the posterior.
\label{fig:regression_lines}}
\end{minipage}
\end{figure}

The output files themselves are typically not immediately useful.
The goal of running
NS is usually to obtain posterior samples and the marginal likelihood $Z$,
whereas {\tt sample.txt} only contains samples from the mixture of constrained
priors. Additional
post-processing is required. This can be achieved by running the following
\proglang{Python} function\footnote{Alternatively, the file
{\tt showresults.py} in the example directory runs this function, and then
calls the code in {\tt display.py} to create a further plot shown in
Figure~\ref{fig:regression_lines}.}:

\begin{CodeChunk}
\begin{CodeInput}
import dnest4
dnest4.postprocess()
\end{CodeInput}
\end{CodeChunk}
This produces the three diagnostic plots in Figures~\ref{fig:fig1},
\ref{fig:fig2}, and~\ref{fig:fig3}, along with the following output:
\begin{CodeChunk}
\begin{CodeOutput}
log(Z) = -175.512920079
Information = 15.1683473499 nats.
Effective sample size = 1521.58236273
\end{CodeOutput}
\end{CodeChunk}
These are the natural log of the marginal likelihood, the
information
\begin{align}
\mathcal{H} &= \int p(\theta|D, M)
\ln\left[\frac{p(\theta | D, M)}{p(\theta | M)}\right] \, d\theta,
\end{align}
which quantifies the degree to which $D$ restricted the
range of possible $\theta$ values,
and the effective sample size,
or number of saved particles with significant posterior weight.
The \code{postprocess} function also saves a file,
{\tt posterior\_sample.txt}, containing posterior samples (one per row).
Unfortunately, it is harder to compute justified error bars on $\ln(Z)$
in DNS than it is in standard NS.
%The \code{postprocess} function does include some options
%for calculating error bars, but they are not used by default,
%and we do not describe them here.
%Hence, the zero-sized error bars on $\ln(Z)$ and $\mathcal{H}$ should
%be ignored.

The \code{postprocess} function can be called while \pkg{DNest4} is running.
This is helpful for monitoring the progress of a run, by inspecting the
output plots.

\subsection{Options}\label{sec:options}
A plain-text file called {\tt OPTIONS} resides in the directory from which you
execute a run. This file contains numerical parameters controlling the
DNS algorithm. Here are the contents of {\tt OPTIONS} for the linear
regression example:

\begin{CodeChunk}
\begin{CodeInput}
# File containing parameters for DNest4
# Put comments at the top, or at the end of the line.
5       # Number of particles
10000   # new level interval
10000   # save interval
100     # threadSteps - pooling interval
0       # maximum number of levels (0 ==> automatic)
10      # Backtracking scale length (lambda in the paper)
100     # Equal weight enforcement. Beta in the paper
10000   # Maximum number of saves (0 ==> run forever)
\end{CodeInput}
\end{CodeChunk}

Additional options are available on the command line. These are
described in Appendix~\ref{sec:command_line_options}.

\subsection{Number of particles}
The first option is the number of particles, here set to five.
If you use more particles, the same amount of CPU time will be spent evolving more particles,
so each one will not be evolved as far. On most problems, five is a sensible
default value. On complex problems where the likelihood function has
a challenging structure, more particles are useful, but it is usually better
to run in multi-threaded mode (see Appendix~\ref{sec:command_line_options}).

\subsection{New level interval}
The new level interval controls how quickly \dnest~creates new levels. In this
example, this is set to 10,000, so a new level will be created once 10,000
MCMC steps have resulted in
10,000 likelihood values above the current top level.
It is difficult to give a sensible default for this
quantity because it depends on the complexity of the problem (basically,
how good the Metropolis proposals are at exploring the target distribution).
However, 10,000 will work for many problems, so we suggest it as a sensible
default. Higher values are slower, but more fail-safe.

\subsection{Save interval}
The save interval controls how often \dnest~writes a model to the output
files; what is usually called ``thinning''. Saving more frequently
(i.e., a smaller save interval) is usually better. However, this can result
in
big output files if your model prints a lot of parameters to {\tt sample.txt}.
causing the postprocessing to take a long time and/or a lot of RAM.
A default suggestion, used in the example, is to set the save interval to the
same value as the new level interval.

\subsection{Thread steps}
The ``thread steps'' parameter controls how frequently separate threads pool
their information about the levels (when running in multi-threaded mode,
see Appendix~\ref{sec:command_line_options}). It should be set to a moderate
value, but should also be a small fraction of the new level interval and the
save interval. 100 is a suggested default value that should work without
problems in the vast majority of cases.

\subsection{Maximum number of levels}
As the name suggests, this tells \pkg{DNest4} how many levels to create, and
therefore controls the factor by which the parameter space is ultimately
compressed. An appropriate value for this quantity depends on the specific
model and dataset at hand --- typically, a larger numbers of parameters
and larger (more informative) datasets will lead to a larger value of
$\mathcal{H}$, and therefore need more levels.

In the initial phase of DNS when levels are being created, the particles
move to the left in Figure~\ref{fig:fig3}, increasing likelihood $L$ and
decreasing prior mass $X$. The posterior distribution is concentrated
where the rate of increase of $\ln(L)$ and the rate of decrease of
$\ln(X)$ are approximately equal.
Figure~\ref{fig:fig3} is the most useful diagnostic
plot for setting the correct
maximum number of levels. A clear peak should be visible in the posterior
weights plot in the lower panel, such that moving further to the left would not
add any more particles with comparable weight to those in the peak.

As described by \citet{skilling2006nested}, there is no guarantee in principle that
another peak might have appeared had a run continued for longer. These
{\em phase changes} are more common in statistical mechanics problems than
in data analysis problems, but can appear in the latter
\citep[e.g.,][]{brewer2014inference, brewer2015fast}.

Alternatively, \pkg{DNest4} can try to determine the maximum number of levels
automatically, if you set the maximum number of levels to 0. This works well
on problems where the MCMC exploration is efficient. When it fails, this
is detectable as the posterior weights plot (lower panel of
Figure~\ref{fig:fig3}) will peak at the left end of its domain. However,
such a failed run may still be useful as it can be used to suggest an order of
magnitude for the required number of levels. For example, if the automatic
setting results in 150 levels and is later seen to fail, it might be worth
a try to set the maximum number of levels to, say, $1.3 \times 150 = 195$.

\subsection{Backtracking scale length}
The backtracking scale length, denoted by $\lambda$, appeared in
Equation~\ref{eqn:weighting}, and controls the degree to which particles
are allowed to ``backtrack'' down in level during the first stage of DNS
when levels are being built. Higher values are more fail-safe, but
make it take longer to create the levels. The value 10 used in the linear
regression example is a suitable default value that should work in almost all
cases. In simple problems where MCMC exploration is easy, lower values
from 1 to 5 work sufficiently well.

\subsection{Equal weight enforcement}
This value is the parameter $\beta$ described in
\citet{brewer2011diffusive}, and compensates
for imprecision in the spacing of the levels, so that the desired mixture
weights $w_j \propto 1$ are achieved during the second stage of the DNS
algorithm. The value 100 is recommended.

\subsection{Maximum number of saves}
This controls the length of a \pkg{DNest4} run, in units of saved particles
in {\tt sample.txt}, which represent the mixture of constrained priors.
The number of posterior samples is always less than this. In most applications
5,000 (as in the linear regression example) provides enough posterior
samples (typically a few hundred) for sufficiently accurate posterior
summaries. However, if you want to plot smooth-looking posterior histograms,
you'll need to increase this value.

If you set the maximum number of saves to zero, \pkg{DNest4} will run until
you terminate it manually.

\section{Implementing models}\label{sec:models}
The ``classic'' method of implementing models in \pkg{DNest4} is by
writing a \proglang{C++} class, an object of which represents a
point in your parameter space.

To run \pkg{DNest4} on any particular problem, such as this linear regression
example, the user needs to define a \proglang{C++} class to specify the
model. Specifically, an object of the class represents a point in the
model's parameter space. Member functions are defined which generate
the object's parameters from the prior, make proposal steps, evaluate the
likelihood, and so on. The sampler calls these member functions while
executing a run.

For the simple linear regression example, we will call the class
{\tt StraightLine}. The member variables representing the
unknown parameters are defined in the header file {\tt StraightLine.h}:

\begin{CodeChunk}
\begin{CodeInput}
class StraightLine
{
    private:
        double m, b, sigma;
};
\end{CodeInput}
\end{CodeChunk}

The class must also define and implement the following member functions:
\begin{enumerate}[(i)]
\item \code{void from\_prior(DNest4::RNG\& rng)}, which generates parameter
        values from the prior;
\item \code{double perturb(DNest4::RNG\& rng)}, which proposes a
        change to the parameter values, and returns $\ln(H)$ as defined in
        Equation~\ref{eqn:logH};
\item \code{double log\_likelihood() const}, which evaluates the log of
        the likelihood function;
\item \code{void print(std::ostream\& out) const}, which prints parameters of
        interest to the given output stream; and
\item \code{std::string description() const}, which returns a \proglang{C++}
        string naming the parameters printed by \code{print(std::ostream\&) const}.
        This string is printed (after a comment character \#) at the top of
        the output file.
\end{enumerate}
These are all described (using the linear regression example) in the
following sections.

%%\subsection{generate}

%\section{Python package}

%% Assign to DFM

%\section{Examples}
% Reasonably simple examples

\subsection{Generating from the prior}
The member function used to generate straight line parameters from the
prior is:
\begin{CodeChunk}
\begin{CodeInput}
void StraightLine::from_prior(DNest4::RNG& rng)
{
   // Naive diffuse prior
   m = 1E3 * rng.randn();
   b = 1E3 * rng.randn();

   // Log-uniform prior
   sigma = exp( - 10.0 + 20.0 * rng.rand());
}
\end{CodeInput}
\end{CodeChunk}

This generates $m$, $b$, and $\sigma$ from their joint prior. In this case,
the priors are all independent, so this reduces to generating the parameters
each from their own prior distribution. For convenience,
\pkg{DNest4} provides an \code{RNG} class to represent random number
generators. The \code{RNG} class is just a convenience wrapper for the
random number generators built into \proglang{C++11}.
As you might expect, there are
\code{rand()} and \code{randn()} member functions to generate
double precision values
from a Uniform$(0,1)$ and Normal$(0,1)$ distribution respectively.

\subsection{Proposal moves}\label{sec:perturb}
The \code{perturb} member function for the straight line model is given below.
This takes a random number generator as input, makes changes to
a subset of the parameters, and returns a \code{double} corresponding
to $\ln(H)$ as defined in Equation~\ref{eqn:logH}. The combination of
the choice of proposal and the $\ln(H)$ value returned must be consistent
with the prior distribution.
\begin{CodeChunk}
\begin{CodeInput}
double StraightLine::perturb(DNest4::RNG& rng)
{
    // log_H value to be returned
    double log_H = 0.0;

    // Proposals must be consistent with the prior
    // Choose which of the three parameters to move
    int which = rng.rand_int(3);

    if(which == 0)
    {
        // log_H takes care of the prior ratio
        // i.e., log_H = log(pi(theta')/pi(theta))
        log_H -= -0.5 * pow(m / 1E3, 2);

        // Take a step
        m += 1E3 * rng.randh();

        // log_H takes care of the prior ratio
        log_H += -0.5 * pow(m / 1E3, 2);
    }
    else if(which == 1)
    {
        log_H -= -0.5*pow(b / 1E3, 2);
        b += 1E3 * rng.randh();
        log_H += -0.5*pow(b / 1E3, 2);
    }
    else
    {
        // Proposal for a parameter with a log-uniform
        // prior takes the log of the parameter,
        // takes a step with respect to a uniform prior,
        // then takes the exp of the parameter
        sigma = log(sigma);
        sigma += 20.0 * rng.randh();

        // Wrap proposed value back into the
        // interval allowed by the prior
        DNest4::wrap(sigma, -10.0, 10.0);
        sigma = exp(sigma);
    }

    return log_H;
}
\end{CodeInput}
\end{CodeChunk}

This function first chooses a random integer from $\{0, 1, 2\}$ using the
\code{rand\_int(int)} member function of the \code{DNest4::RNG} class.
This determines which of the three parameters
$(m, b, \sigma)$ is modified. In this example, there are no proposals
that modify more than one of the parameters at a time, and all proposals
are ``random walk'' proposals that add a perturbation
(drawn from a symmetric distribution) to the current value.

\subsection{Proposals for single parameters}\label{sec:proposals}
The proposal for $m$ involves adding a perturbation to the current
value using the line

\begin{CodeChunk}
\begin{CodeInput}
m += 1E3 * rng.randh();
\end{CodeInput}
\end{CodeChunk}

A challenge using MCMC for standard Nested Sampling is that the target
distribution is not static --- it gets compressed over time. Similarly, in
the first stage of DNS the target distribution gets compressed (as levels are
added), and in the second stage the target distribution is a mixture of
distributions that have been compressed to varying degrees.
This makes it difficult to tune step-sizes as you would when using
the standard Metropolis algorithm to sample the posterior distribution.

Rather than trying to adapt proposal distributions as a function of level,
it is much simpler to just use heavy-tailed proposals which have some
probability of making a jump of appropriate size. This is slightly
wasteful of CPU time, but it saves a lot of human time and is more
fail-safe than tuned step sizes.
In simple experiments, we have found that heavy-tailed proposals are
about as efficient as slice sampling \citep{neal2003slice}, but much easier to
implement. The following procedure generates
$x$ from a heavy-tailed distribution:

\begin{enumerate}
\item Generate $a \sim$ Normal$(0, 1)$;
\item Generate $b \sim$ Uniform$(0, 1)$;
\item Define $t := a/\sqrt{-\ln(b)}$;
\item Generate $n \sim$ Normal$(0, 1)$;
\item Set $x := 10^{1.5 - 3|t|}n$
\end{enumerate}
The variable $t$ has a student-$t$ distribution with 2 degrees of freedom.
Overall, this procedure generates values $x$
with a maximum scale of tens--hundreds, down to
a minimum scale of about $10^{-30}$ with 99\% probability, by virtue
of the $t$-distribution's heavy tails.
For convenience, the \code{RNG} class contains a member function \code{randh()}
to generate values from this distribution.
The factor of \code{1E3} is included because it is a measure of the prior
width. Therefore, this proposal will attempt moves whose maximum order of
magnitude is a bit wider than the prior (since it would be very surprising
if any bigger moves were needed), and will propose moves a few orders of
magnitude smaller with moderate probability. We recommend using
this strategy (a measure of prior width, multiplied by \code{randh()}) as
a default proposal that works well in almost all problems.

Recall that for Nested Sampling, the Metropolis acceptance probability,
excluding the term for the likelihood, is
\begin{align}
H = \frac{q(\params'|\params)}{q(\params | \params')}
\times \frac{\pi(\params')}{\pi(\params)}.
\end{align}
When implementing a model class, the
\code{perturb(DNest4::RNG\&)}
member function must return the logarithm of this value.
Since the prior for $m$ is a normal distribution with
mean zero and standard deviation $1000$, $H$ is
\begin{align}
H &= \frac{\exp\left[-\frac{1}{2}(m'/1000)^2\right]}
{\exp\left[-\frac{1}{2}(m/1000)^2\right]}.
\end{align}
This explains the use of the \code{log\_H} variable.

In the example, the proposal for $\sigma$ is implemented by taking advantage
of the uniform prior for $\ln(\sigma)$. So $\sigma$ is transformed by
taking a logarithm, a proposal move is made (that satisfies detailed balance
with respect to a uniform distribution), and then $\sigma$ is exponentiated
again. The step for the uniform prior between -10 and +10
uses the following code:
\begin{CodeChunk}
\begin{CodeInput}
sigma += 20.0 * rng.randh();
DNest4::wrap(sigma, -10.0, 10.0);
\end{CodeInput}
\end{CodeChunk}
The factor of 20 accounts for the prior width, and
the \code{wrap(double\&, double, double)} function uses the modulo operator
to construct periodic boundaries. For example, if the perturbation results
in a value of 10.2, which is outside the prior range, the value is modified to
$-9.8$.
The \code{wrap} function has no return value and works by modifying its
first argument, which is passed by reference.
Alternatively, the log-uniform prior, which has density proportional
to $1/\sigma$, could have been used directly by adding
$\ln\left[(1/\sigma')/(1/\sigma)\right]$ to the
return value instead of using the log/exp trick.
However, this isn't recommended for a prior distribution like this
which covers several orders of magnitude, because the appropriate scale size
for the proposal is less clear. This is likely to cause inefficient
sampling.

\subsection{Consistency of prior and proposal}
It is imperative that \code{from_prior} and
\code{perturb}
be consistent with each other, and that each implements
the prior distributions that you want to use. One technique
for testing this is to sample the prior for a long time
(by setting the maximum number of levels to 1) and inspect
{\tt sample.txt} to ensure that each parameter is exploring
the prior correctly.

\subsection{Log likelihood}
The log-likelihood for the model is evaluated and returned by the
\code{log\_likelihood} member function. For the straight line
fitting example, the log likelihood is based on the
normal density, and the code is given below.
\begin{CodeChunk}
\begin{CodeInput}
double StraightLine::log_likelihood() const
{
    // Grab the dataset
    const std::vector<double>& x = Data::get_instance().get_x();
    const std::vector<double>& y = Data::get_instance().get_y();

    // Variance
    double var = sigma * sigma;

    // Conventional gaussian sampling distribution
    double log_L = 0.0;
    double mu;
    for(size_t i = 0; i < y.size(); ++i)
    {
        mu = m * x[i] + b;
        log_L += -0.5 * log(2 * M_PI * var) - 0.5 * pow(y[i] - mu, 2) / var;
    }

    return log_L;
}
\end{CodeInput}
\end{CodeChunk}
The dataset is assumed to be accessible inside this function. In the
regression example, this is achieved by having a
\code{Data} class to represent
datasets. Since there will usually only be one dataset,
the {\em singleton pattern} (a class of which there is one instance
accessible from anywhere) is recommended.
The \code{Data} class has
one static member which is itself an object of class \code{Data}, and is
accessible using \code{Data::get_instance()} --- this is the singleton pattern,
essentially a way of defining quasi-`global' variables:

\begin{CodeChunk}
\begin{CodeInput}
class Data
{
    private:
        // 'static' means this exists at the level of the class
        // instead of being truly global
        static Data instance;

    public:
        // Getter
        static Data& get_instance();
};
\end{CodeInput}
\end{CodeChunk}

Alternatively, the data could be defined using static members of your
model class.

\subsection{Parameter output}
The print and description functions are very simple:
\begin{CodeChunk}
\begin{CodeInput}
void StraightLine::print(std::ostream& out) const
{
    out << m << ' ' << b << ' ' << sigma;
}

std::string StraightLine::description() const
{
    return std::string("m, b, sigma");
}
\end{CodeInput}
\end{CodeChunk}
The print function specifies that the parameters
$m$, $b$, and $\sigma$ are printed a single line
(of {\tt sample.txt}) and are separated by spaces (the
\code{postprocess} function assumes the delimeter is a space).

\subsection{Running the sampler}
The file {\tt main.cpp} contains the \code{main()} function which
is executed after compiling and linking. The contents of {\tt main.cpp}
for the linear regression example are:

\begin{CodeChunk}
\begin{CodeInput}
#include <iostream>
#include "Data.h"
#include "DNest4/code/DNest4.h"
#include "StraightLine.h"

using namespace std;

int main(int argc, char** argv)
{
    Data::get_instance().load("road.txt");
    DNest4::start<StraightLine>(argc, argv);
    return 0;
}
\end{CodeInput}
\end{CodeChunk}

The first line of {\tt main()} loads the data into its global instance
(so it can be accessed from within the log-likelihood function)
and the second line uses a \code{start} template function to construct
and run the sampler.

\section[Finite mixture models with the RJObject class]{Finite mixture models with the \code{RJObject} class}
Mixture models are a useful way of representing realistic prior information
in Bayesian data analysis. To reduce the amount of effort needed to
implement mixture models in DNS, \citet{brewer2014inference} implemented
a template class called \code{RJObject} to handle the MCMC moves
required. The RJ in \code{RJObject} stands for Reversible Jump
\citep{green1995reversible}, as
\code{RJObject} implements birth and death moves for mixture
models with an unknown number of components. An updated version of
\code{RJObject} is included in \pkg{DNest4}.

\begin{figure}[ht!]
\begin{center}
\hspace*{-2cm}\includegraphics[width=0.4\textwidth]{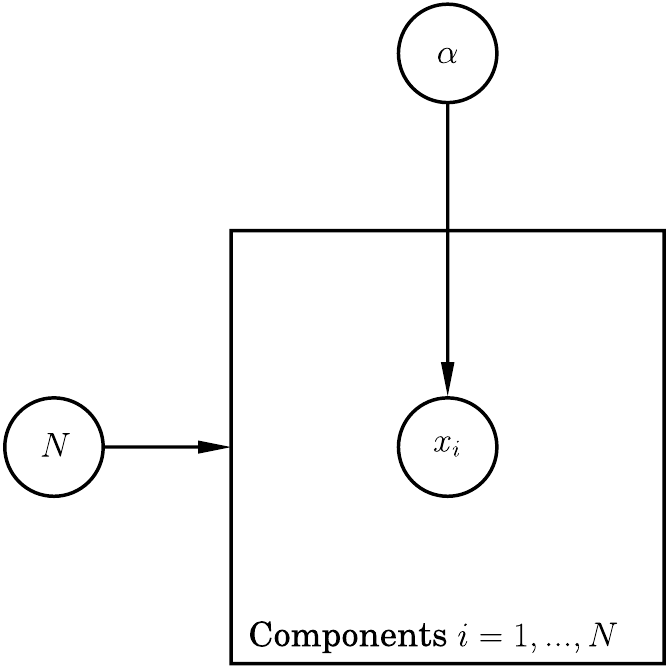}
\caption{A PGM showing the kind of prior information the \code{RJObject}
template class expresses.
Figure created using Daft (\url{http://daft-pgm.org}).\label{fig:rjobject_pgm}}
\end{center}
\end{figure}

If $N$ is the number of components, $x_i$ denotes the vector of parameters
of the $i$th component, and $\alpha$ is the vector of hyperparameters, the prior
can be factorized via the product rule, giving
\begin{align}
p\left(N, \alpha, \left\{x_i\right\}_{i=1}^N\right)
&= p(N)p(\alpha | N)p\left(\left\{x_i\right\}_{i=1}^N | \alpha, N\right).
\end{align}
The specific assumptions of \pkg{RJObject} are that this simplifies to
\begin{align}
p\left(N, \alpha, \left\{x_i\right\}_{i=1}^N\right)
&= p(N)p(\alpha)
\prod_{i=1}^N
p\left(x_i | \alpha\right).\label{eqn:rjobject}
\end{align}
That is, the prior for the hyperparameters is independent of the number of
components, and each component is independent and identically
distributed given the hyperparameters. The sampling distribution
$p(D | x, \alpha)$ (not shown in the PGM)
cannot depend on the ordering of the components.
A probabilistic graphical model (PGM) is showing this dependence
structure is shown in Figure~\ref{fig:rjobject_pgm}. There are no observed
data nodes here, but if such a structure forms part of a
Bayesian model, an \code{RJObject} object within your model class
can encapsulate this part of the model.

\citet{brewer2014inference}
described the motivation for \code{RJObject} and some details
about the Metropolis proposals underlying it. Here, we demonstrate how to
implement a finite mixture model using \code{RJObject}.
This example can be found in the
{\tt code/Examples/RJObject\_1DMixture}
directory. The ``SineWave'' and ``GalaxyField'' models
from \citet{brewer2014inference}
are also included in the \pkg{DNest4} repository.

\subsection{An example mixture model}
Consider a sampling distribution
for data $D=\{D_1, D_2, ..., D_n\}$ which is a mixture of $N$ gaussians
with means $\{\mu_j\}$, standard deviations $\{\sigma_j\}$, and
mixture weights $\{w_j\}$. The likelihood for the $i$th data point is
\begin{align}
p\left(D_i | N, \{\mu_j\}, \{\sigma_j\}, \{w_j\}\right) &=
\sum_{j=1}^N \frac{w_j}{\sigma_j\sqrt{2\pi}}
\exp\left[-\frac{1}{2\sigma_j^2}\left(D_i - \mu_j\right)^2\right].
\end{align}
Colloquially, one might say the data were `drawn from' a mixture of
$N$ normal distributions, and we want to infer $N$ along with the
properties (positions, widths, and weights) of those normal distributions.
The mixture weights $\{w_j\}$ must obey a normalization condition
$\sum_j w_j = 1$. The easiest way of implementing this is to use
un-normalized weights $\{W_j\}$ which do not obey
such a condition and then normalize them by dividing by their sum.

Making the connection with Equation~\ref{eqn:rjobject} and
Figure~\ref{fig:rjobject_pgm}, the ``components'' are the gaussians,
the gaussian parameters are $\{x_j\} = \{(\mu_j, \sigma_j, W_j)\}$,
and hyperparameters $\alpha$ may be used to help specify
a sensible joint prior for the gaussian parameters.

We now describe the specific prior distributions we used in this example.
The prior for $N$ was
\begin{align}
p(N) \propto \frac{1}{N+1}
\end{align}
for $N \in \{1, 2, ..., 100\}$.
For the conditional prior of the gaussian parameters,
we used Laplace (biexponential)
distributions\footnote{A Laplace
distribution with location parameter $a$ and scale parameter
$b$ has density $p(x | a, b) = \frac{1}{2}\exp\left(-\frac{1}{b}|x - a|\right)$.}.
Normal distributions would be more convenional, but the analytic cumulative
distribution function (CDF) of the Laplace distribution makes it easier to
implement. These were:
\begin{align}
\mu_j &\sim \textnormal{Laplace}(a_\mu, b_\mu)\label{eqn:cond_prior1}\\
\ln\sigma_j &\sim \textnormal{Laplace}(a_{\ln\sigma}, b_{\ln\sigma})\label{eqn:cond_prior2}\\
\ln W_j &\sim \textnormal{Laplace}(0, b_{\ln W}).\label{eqn:cond_prior3}
\end{align}
The location parameters are denoted with $a$ and scale parameters with $b$.
These priors express the idea that the centers, widths, and relative weights
of the gaussian mixture components are probably clustered around some typical
value.

We used the following priors for the hyperparameters:
\begin{align}
a_\mu &\sim \textnormal{Uniform}(-1000, 1000)\\
\ln b_\mu &\sim \textnormal{Uniform}(-10, 10)\\
a_{\ln\sigma} &\sim \textnormal{Uniform}(-10, 10)\\
b_{\ln\sigma} &\sim \textnormal{Uniform}(0, 5)\\
b_{\ln W} &\sim \textnormal{Uniform}(0, 5).
\end{align}
The normalized weights are $w_k = W_k/\sum w_j$.

\subsection{Using the RJObject class}
To implement this model for \pkg{DNest4}, the model
class needs to contain an instance of an
\code{RJObject}, which contains the parameters
$\{(\mu_j, \sigma_j, w_j)\}$:
\begin{CodeChunk}
\begin{CodeInput}
class MyModel
{
    private:
        DNest4::RJObject<MyConditionalPrior> gaussians;
....
\end{CodeInput}
\end{CodeChunk}
where the template argument \code{<MyConditionalPrior>}
is a class implementing the prior for the hyperparameters
$\alpha$ and the form of the conditional prior $p(x_i | \alpha)$.
The main advantage of the \code{RJObject} class is that
we do not need to implement any proposals for $\{(\mu_j, \sigma_j, W_j)\}$.
Rather, these can be done trivially as follows:
\begin{CodeChunk}
\begin{CodeInput}
    // Generate the components from the prior
    gaussians.from_prior(rng);

    // Do a Metropolis proposal
    double logH = gaussians.perturb(rng);

    // Print to output stream 'out'
    gaussians.print(out);
\end{CodeInput}
\end{CodeChunk}

The \code{RJObject} constructor definition is:
\begin{CodeChunk}
\begin{CodeInput}
RJObject(int num_dimensions, int max_num_components, bool fixed,
         const ConditionalPrior& conditional_prior,
         PriorType prior_type=PriorType::uniform);
\end{CodeInput}
\end{CodeChunk}
where \code{num\_dimensions} is the number of parameters needed to
specify a single component (three in this example), \code{max\_num\_components}
is the maximum value of $N$ allowed, \code{fixed} determines whether
$N$ is fixed at $N_{\rm max}$ or allowed to vary,
\code{conditional\_prior} is an instance of a conditional prior,
and \code{prior\_type} controls the prior for $N$ (the default is uniform
from $\{0, 1, 2, ..., N_{\rm max}\}$). The {\tt RJObject\_1DMixture}
\code{MyModel} initializes its {\tt RJObject} as follows:
\begin{CodeChunk}
\begin{CodeInput}
MyModel::MyModel()
:gaussians(3, 100, false, MyConditionalPrior(), PriorType::log_uniform)
{
}
\end{CodeInput}
\end{CodeChunk}
Passing \code{PriorType::log_uniform} for the final argument
specifies the $1/(N+1)$ prior for $N$.
One complication for this model is that \code{RJObject}, by default,
allows $N$ to be zero, which makes no sense for this particular
problem. Therefore, we prevent $N=0$ from being generated in
\code{MyModel::from\_prior}, and assert that it should always be
rejected if proposed in \code{MyModel::perturb}:

\begin{CodeChunk}
\begin{CodeInput}
void MyModel::from_prior(RNG& rng)
{
    do
    {
        gaussians.from_prior(rng);
    }while(gaussians.get_components().size() == 0);
}

double MyModel::perturb(RNG& rng)
{
    double logH = 0.0;

    logH += gaussians.perturb(rng);
    if(gaussians.get_components().size() == 0)
        return -std::numeric_limits<double>::max();

    return logH;
}
\end{CodeInput}
\end{CodeChunk}

To access the component parameters, the \code{RJObject} member function
\code{get_components} is used. This was used in the above functions, but more
typically it is needed in the log likelihood.
The member function \code{get_components} returns (by const reference) a
\code{std::vector} of \code{std::vector}s of \code{double}s. For example,
parameter two of component zero is:
\begin{CodeChunk}
\begin{CodeInput}
gaussians.get_components()[0][2]
\end{CodeInput}
\end{CodeChunk}
The order of $\{(\mu_j, \sigma_j, w_j)\}$
(i.e., which one is parameter 0, 1, and 2)
is determined by the conditional prior class, and is
explained in the following section.

\code{RJObject}'s \code{print} function prints the dimensionality of each
component and the maximum value of $N$, followed by the hyperparameters,
then parameter zero of each component (zero padded when $N<N_{\rm max}$),
parameter one for each component, and so on.

\subsection{Conditional priors}
The \code{RJObject} class is used to manage the components. An additional
class is needed to define and manage the hyperparameters $\alpha$ (i.e.,
define how they are generated from the prior, how they are proposed, and
so on). The iid
conditional prior $p(x_i | \alpha)$ is also defined by this extra class.
In the example, the class is called \code{MyConditionalPrior} and
is inherited from an abstract base class \code{DNest4::ConditionalPrior}
which insists that certain member functions be specified.

The member functions \code{from\_prior}, \code{perturb\_hyperparameters}, and
\code{print} do the same things as the similar functions in a standard model
class, but for the hyperparameters $\alpha$. 
In addition, three functions
\code{from\_uniform}, \code{to\_uniform}, and \code{log\_pdf}
together define the form of the conditional prior for the
component parameters, $p(x_i | \alpha)$. Each of these takes a
\code{std::vector} of \code{double}s by reference as input.

The \code{log\_pdf} function just evaluates $\ln p(x_i | \alpha)$
at the current value of $\alpha$. The position $x_i$ where this is evaluated
is passed in via the input vector. In the example, these
were defined using Equations~\ref{eqn:cond_prior1}--\ref{eqn:cond_prior3}.
To simplify this, we have defined a class for Laplace distributions,
rather than explicitly writing out the densities. We decided (arbitarily)
that parameters 0,1, and 2 are $\mu$, $\ln(\sigma)$, and $W$ respectively.
\begin{CodeChunk}
\begin{CodeInput}
// vec = {mu, log_sigma, log_weight}
double MyConditionalPrior::log_pdf(const std::vector<double>& vec) const
{
    // Three Laplace distributions
    Laplace l1(location_mu, scale_mu);
    Laplace l2(location_log_sigma, scale_log_sigma);
    Laplace l3(0.0, scale_log_weight);
    return l1.log_pdf(vec[0]) + l2.log_pdf(vec[1]) + l3.log_pdf(vec[2]);
}
\end{CodeInput}
\end{CodeChunk}

The functions \code{to\_uniform} and \code{from\_uniform} must implement
the cumulative distribution function (CDF) of the conditional
prior and its inverse. i.e., \code{from\_uniform}, if the input vector contains
iid draws from Uniform(0, 1), these should be modified to become draws from
$p(x|\alpha)$, and \code{to\_uniform} should be the inverse of
\code{from\_uniform}.
Both of these functions modify the vector argument in-place.
For the example, we again used an external class to define these functions
for the Laplace distribution:

\begin{CodeChunk}
\begin{CodeInput}
// vec (input) = {u1, u2, u3} ~ Uniform(0, 1) in the prior
// gets modified to {mu, log_sigma, log_weight}
void MyConditionalPrior::from_uniform(std::vector<double>& vec) const
{
    // Three Laplace distributions
    Laplace l1(location_mu, scale_mu);
    Laplace l2(location_log_sigma, scale_log_sigma);
    Laplace l3(0.0, scale_log_weight);

    vec[0] = l1.cdf_inverse(vec[0]);
    vec[1] = l2.cdf_inverse(vec[1]);
    vec[2] = l3.cdf_inverse(vec[2]);
}

// vec (input) = {mu, log_sigma, log_weight}
// gets modified to {u1, u2, u3} using the CDF of the conditional prior
// This is the inverse of from_uniform
void MyConditionalPrior::to_uniform(std::vector<double>& vec) const
{
    // Three Laplace distributions
    Laplace l1(location_mu, scale_mu);
    Laplace l2(location_log_sigma, scale_log_sigma);
    Laplace l3(0.0, scale_log_weight);

    vec[0] = l1.cdf(vec[0]);
    vec[1] = l2.cdf(vec[1]);
    vec[2] = l3.cdf(vec[2]);
}
\end{CodeInput}
\end{CodeChunk}

\subsection{Mixture model results}
\begin{figure}
\begin{minipage}{0.45\textwidth}
\centering
\includegraphics[width=0.95\textwidth]{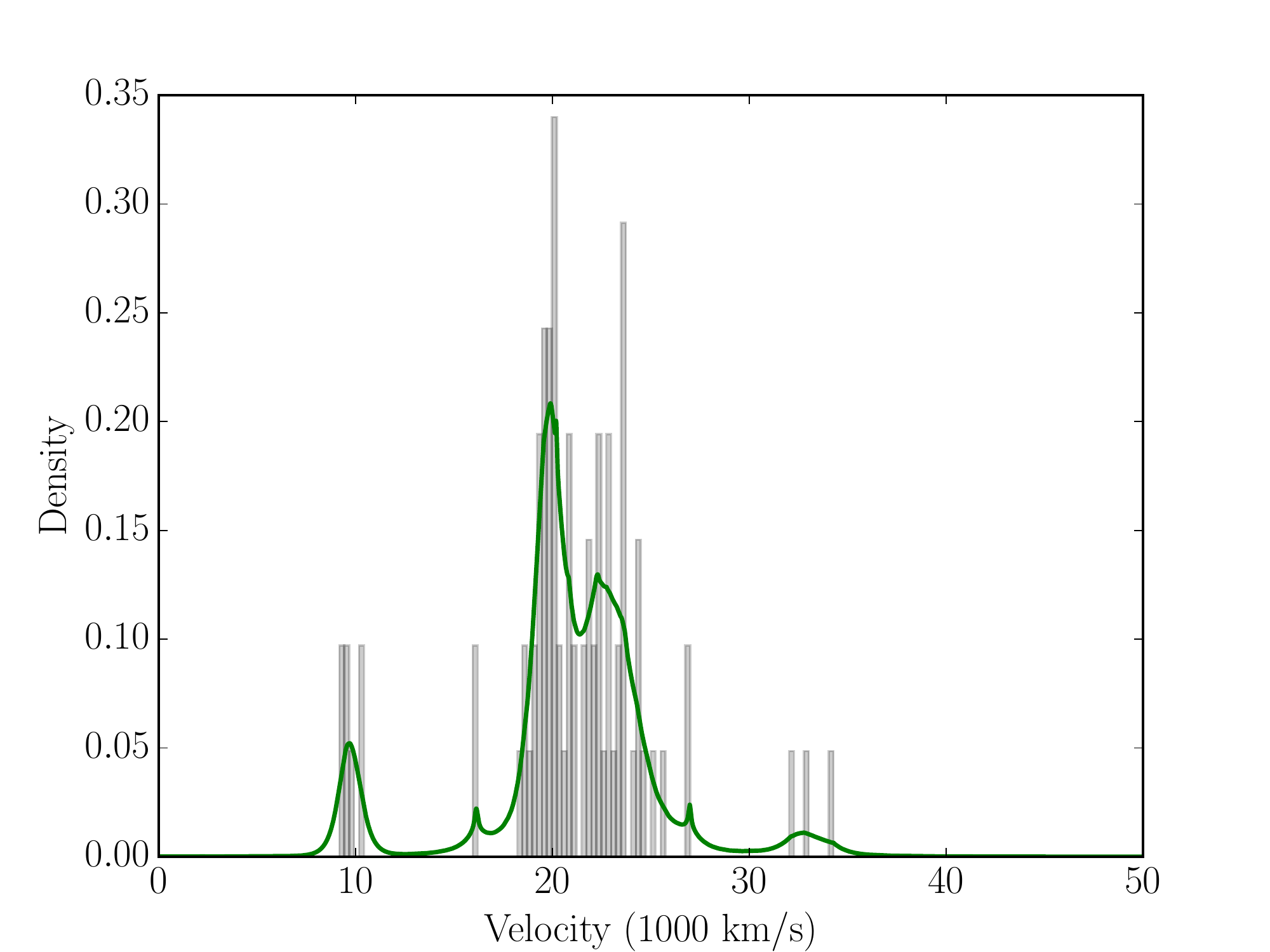}
\caption{The `galaxy' data, with the posterior mean fit
(equivalent to the predictive distribution for the ``next'' data point).\label{fig:galaxies}}
\end{minipage} \hspace{0.05\textwidth}
\begin{minipage}{0.45\textwidth}
\centering
\includegraphics[width=0.95\textwidth]{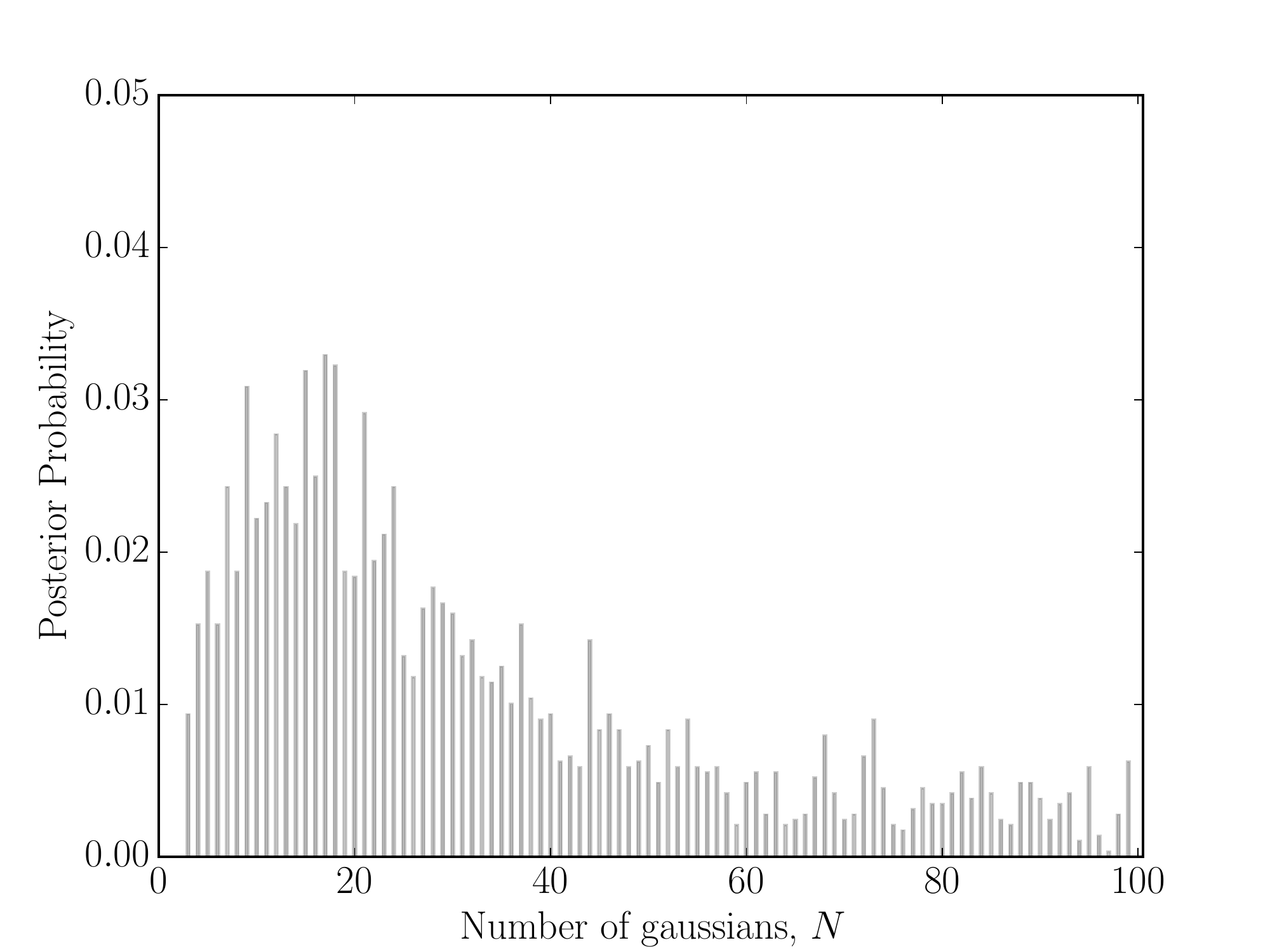}
\caption{The posterior distribution for $N$ given the
galaxy data.\label{fig:galaxies_N}}
\end{minipage}
\end{figure}

The data and the posterior mean fit are shown
in Figure~\ref{fig:galaxies} and the posterior distribution for
$N$ is shown in Figure~\ref{fig:galaxies_N}.
The marginal likelihood was $\ln(Z) = -232.2$, and
the information (measured in nats) was $\mathcal{H} = 29.5$.

\begin{figure}[ht!]
\centering
\includegraphics[width=0.7\textwidth]{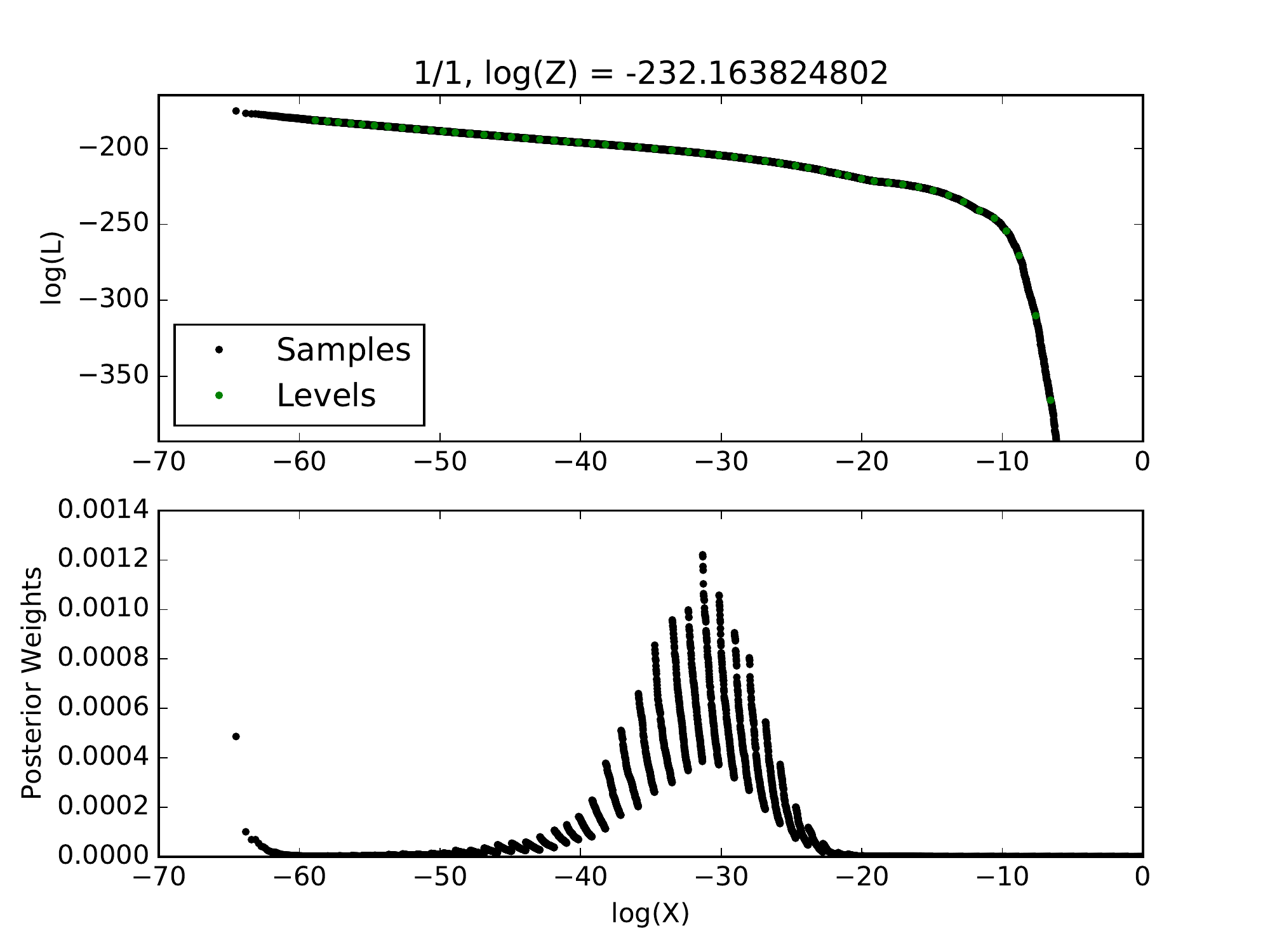}
\caption{The log-likelihood and posterior weights as a function
of compression $X$ for the galaxy data. There are phase changes
\citep{skilling2006nested} for which Nested Sampling is the best known
solution. There are some points with high posterior weight at the left
of the plot, but there are so few of them that they make up a trivial
fraction of the posterior mass. In practice, however, it is worth
re-running with more levels to verify that a second peak is not forming.
\label{fig:galaxies_fig3}}
\end{figure}

\section{Approximate Bayesian computation}
\subsection{Background}
Nested Sampling can be used to solve Approximate Bayesian Computation (ABC)
problems elegantly and efficiently.
Also known (misleadingly) as likelihood-free inference,
ABC
is a set of Monte Carlo techniques for approximating the posterior distribution
without having to evaluate the likelihood function $L(\theta)$. Instead, 
the user must be able to cheaply
generate simulated data sets from the sampling distribution $p(D|\theta, M)$.
Since a sampling distribution must be specified, it is not that there is
no likelihood function (indeed, a sampling distribution and
a data set imply a particular likelihood function), rather that we cannot
evaluate it cheaply and therefore cannot use MCMC.

All Bayesian updating conditions on the truth of a proposition. We sometimes
speak and use notation as if we're conditioning on the value of a variable, for
example by writing the posterior distribution as
$p(\theta | D, M)$. However,
this is shorthand for $p(\theta | D = D_{\rm observed}, M)$.
In the prior state of knowledge, the statement
$D = D_{\rm observed}$ could have been either
true or false, but it is known to be true
in the posterior state of knowledge.
In the case of ``continuous data'' (really a continuous {\it space of
possibilities for the data before we learned it}) we condition on a
proposition like $(D \in \mathcal{R})$ where $\mathcal{R}$ is a region, and
then implicitly take a limit as the size of $\mathcal{R}$ goes to zero.

A simple ``rejection sampling'' version of
ABC works by sampling the joint prior distribution for the parameters and
data
\begin{align}
p(\theta, D | M) &= p(\theta | M)p(D | \theta, M)
\end{align}
and rejecting samples for which $D \neq D_{\rm observed}$, so that the
samples represent
\begin{align}
p(\theta, D | D = D_{\rm observed}, M) &\propto p(\theta | M)p(D | \theta, M)
\mathds{1}\left[D = D_{\rm observed}\right].\label{eqn:abc_posterior}
\end{align}
The marginal distribution for $\theta$ is also the conditional distribution
$p(\theta | D = D_{\rm observed}, M)$, that is,
the posterior\footnote{Incidentally, this usage of the {\em joint} posterior
distribution for the parameters {\em and the data} provides a bridge
connecting Bayesian updating to maximum entropy updating
\citep{caticha2006updating,giffin2007updating}.}.

This approach is rarely usable in practice because the probability of
generating a dataset that matches the observed one is extremely low.
To work around this, we replace the proposition
$D = D_{\rm observed}$ with a logically
weaker one (i.e., one that is implied by $D = D_{\rm observed}$ but does not
imply it).
The weaker proposition is defined using
a discrepancy function
$\rho$, which measures how different a simulated dataset $D$ is from
the real one $D_{\rm observed}$:
\begin{align}
\rho\left(D; D_{\rm observed}\right).
\end{align}
The discrepancy function should take a minimum value of zero when
$D = D_{\rm observed}$.
The analysis then proceeds by Monte Carlo sampling the
joint posterior for $\theta$ and $D$
conditional on the logically weaker proposition
\begin{align}
\rho\left(D; D_{\rm observed}\right) < \epsilon
\end{align}
where $\epsilon$ is some small number. With this proposition, the
rejection rate will still
typically be very high, but lower than with $D = D_{\rm observed}$.
Typically, $\rho$ is defined by introducing a few {\em summary statistics}
$s_1(D), s_2(D), ..., s_n(D)$, which hopefully capture most of the relevant
information in the data, and then computing the Euclidian distance
between the summary statistics of $D$ and $D_{\rm observed}$.
That is,
\begin{align}
\rho(D; D_{\rm observed}) &=
\sqrt{\sum_i \left[s_i(D) - s_i(D_{\rm observed})\right]^2}.
\end{align}

The main challenges associated with ABC are:
\begin{enumerate}[(i)]
\item The choice of discrepancy function $\rho(D; D_{\rm observed})$, which
may involve a choice of summary statistics;
\item How to choose the value of $\epsilon$, or how to change it as a run
progresses; and
\item How to make algorithms more efficient than rejection sampling.
\end{enumerate}
Challenge (i) is Bayesian in nature, i.e., it relates to the very
definition of the posterior distribution itself.
On the other hand, challenges (ii) and (iii) are about the computational
implementation. Most ABC analyses are done using Sequential Monte Carlo
\citep[SMC; ][]{del2012adaptive}, another family of algorithms closely related
to Nested Sampling and annealing in the sense that they work with
sequences of probability distributions.

\subsection{ABC with nested sampling}
NS can be used for ABC by noting that Equation~\ref{eqn:abc_posterior}
is of the same form as a constrained prior, only instead of being the prior
for the parameters $p(\theta | M)$ defined on the parameter space, it is
the {\em joint prior} for the parameters and the data,
$p(\theta, D |M) = p(\theta | M)p(D | \theta, M)$, defined on the product
space of possible parameters and datasets. Therefore, to implement
ABC in \pkg{DNest4}, you implement a model class
whose \code{from\_prior} member function generates parameters {\em and simulated
data} from the joint prior, and whose \code{perturb} member function
proposes changes to the parameters and/or the simulated data, in a
manner consistent with the joint prior. The \code{log\_likelihood} function
of the class, instead of actually evaluating the log likelihood, should
evaluate $-\rho(D; D_{\rm observed})$. Then, running the sampler
will create levels which systematically decrease $\rho$. The samples from
a particular level then represent the ABC posterior defined in
Equation~\ref{eqn:abc_posterior} with $\epsilon$ corresponding to minus
the ``log likelihood'' of the level.
A single \pkg{DNest4} run allows you to test sensitivity
to the value of $\epsilon$ using a single run.
The marginal likelihood reported will be
the probability that $\rho\left(D; D_{\rm observed}\right) < \epsilon$
given the model, although caution should be applied if
using this for model selection \citep{robert2011lack}.

\subsection{ABC example}
A simple ABC example is included in the directory
{\tt code/Examples/ABC}. The example tries to infer the mean
$\mu$ and standard deviation $\sigma$ of a normal distribution
from a set of samples. The sampling distribution is
\begin{align}
x_i &\sim \textnormal{Normal}(\mu, \sigma^2).
\end{align}
and the prior is
\begin{align}
\mu &\sim \textnormal{Uniform}(-10, 10)\\
\ln\sigma &\sim \textnormal{Uniform}(-10, 10).
\end{align}
However, instead of conditioning on the full dataset $\{x_i\}$, the example
uses the minimum and maximum values in the data,
$\min(\{x_i\})$ and $\max(\{x_i\})$, as summary statistics\footnote{This is
not a serious application of ABC, since the likelihood function can easily
be evaluated. It is included to demonstrate the principle of using
Nested Sampling as an ABC method.}.

In ABC applications the model class needs to describe a point in the joint
(parameters, data) space, and the \code{from\_prior} and \code{perturb}
functions need to respect the joint prior distribution.
In Bayesian inference the joint prior tends to make the
parameters and data highly dependent
--- otherwise the data wouldn't ever be informative about the parameters.
To improve efficiency, it is helpful to use a change of variables such that
the parameters and data-controlling variables are independent. In this
example we use

\begin{align}
\mu &\sim \textnormal{Uniform}(-10, 10)\\
\ln\sigma &\sim \textnormal{Uniform}(-10, 10)\\
n_i &\sim \textnormal{Normal}(0, 1)
\end{align}
where each of the $n_i$ enables a data point to be calculated using
\begin{align}
x_i &:= \mu + \sigma n_i.
\end{align}
A general procedure for achieving this is to imagine generating a
simulated data set from the parameters. In this process, you would need
to call a random number generator many times.
The results of the random number calls
are independent of the parameters and
can be used to parameterize the data set. In the example,
the $\{n_i\}$ variables play this role.
Therefore, the variables in the model class are:
\begin{CodeChunk}
\begin{CodeInput}
class MyModel
{
    private:
        double mu, log_sigma;
        std::vector<double> n;
\end{CodeInput}
\end{CodeChunk}

The proposal involves changing either \code{mu}, \code{log\_sigma}, or
one of the \code{n}s:
\begin{CodeChunk}
\begin{CodeInput}
double MyModel::perturb(RNG& rng)
{
    int which = rng.rand_int(3);

    if(which == 0)
    {
        mu += 20 * rng.randh();
        wrap(mu, -10.0, 10.0);
    }
    if(which == 1)
    {
        log_sigma += 20 * rng.randh();
        wrap(log_sigma, -10.0, 10.0);
    }
    if(which == 2)
    {
        int i = rng.rand_int(n.size());
        n[i] = rng.randn();
    }

    return 0.0;
}
\end{CodeInput}
\end{CodeChunk}

The ``log-likelihood'' is really minus the discrepancy function.
Since we parameterized joint (parameter, data) space using the
$n$ variables instead of the data $\{x_i\}$ itself, we must
``assemble'' the simulated dataset in order to evaluate the
discrepancy function:
\begin{CodeChunk}
\begin{CodeInput}
double MyModel::log_likelihood() const
{
    double x_min = Data::get_instance().get_x_min();
    double x_max = Data::get_instance().get_x_max();

    double sigma = exp(log_sigma);

    // Assemble fake dataset
    vector<double> x_fake = n;
    for(size_t i = 0; i < x_fake.size(); i++)
        x_fake[i] = mu + sigma * x_fake[i];

    // Goodness
    double logL = 0.;
    logL -= pow(*min_element(x_fake.begin(), x_fake.end()) - x_min, 2);
    logL -= pow(*max_element(x_fake.begin(), x_fake.end()) - x_max, 2);

    return logL;
}
\end{CodeInput}
\end{CodeChunk}

\subsection{ABC example results}
For ABC applications, there is an alternate \code{postprocess} function
called \code{postprocess_abc}.
This generates posterior samples using the $\epsilon$ value corresponding
to level $\approx$ 3/4$\times$ (maximum number of levels).
\begin{CodeChunk}
\begin{CodeInput}
dnest4.postprocess_abc()
\end{CodeInput}
\end{CodeChunk}
To try different values of $\epsilon$, you can set the argument
\code{threshold\_fraction} to a value other than 0.8. For example
\begin{CodeChunk}
\begin{CodeInput}
dnest4.postprocess_abc(threshold_fraction = 0.6)
\end{CodeInput}
\end{CodeChunk}

\begin{figure}[ht!]
\centering
\includegraphics[width=0.6\textwidth]{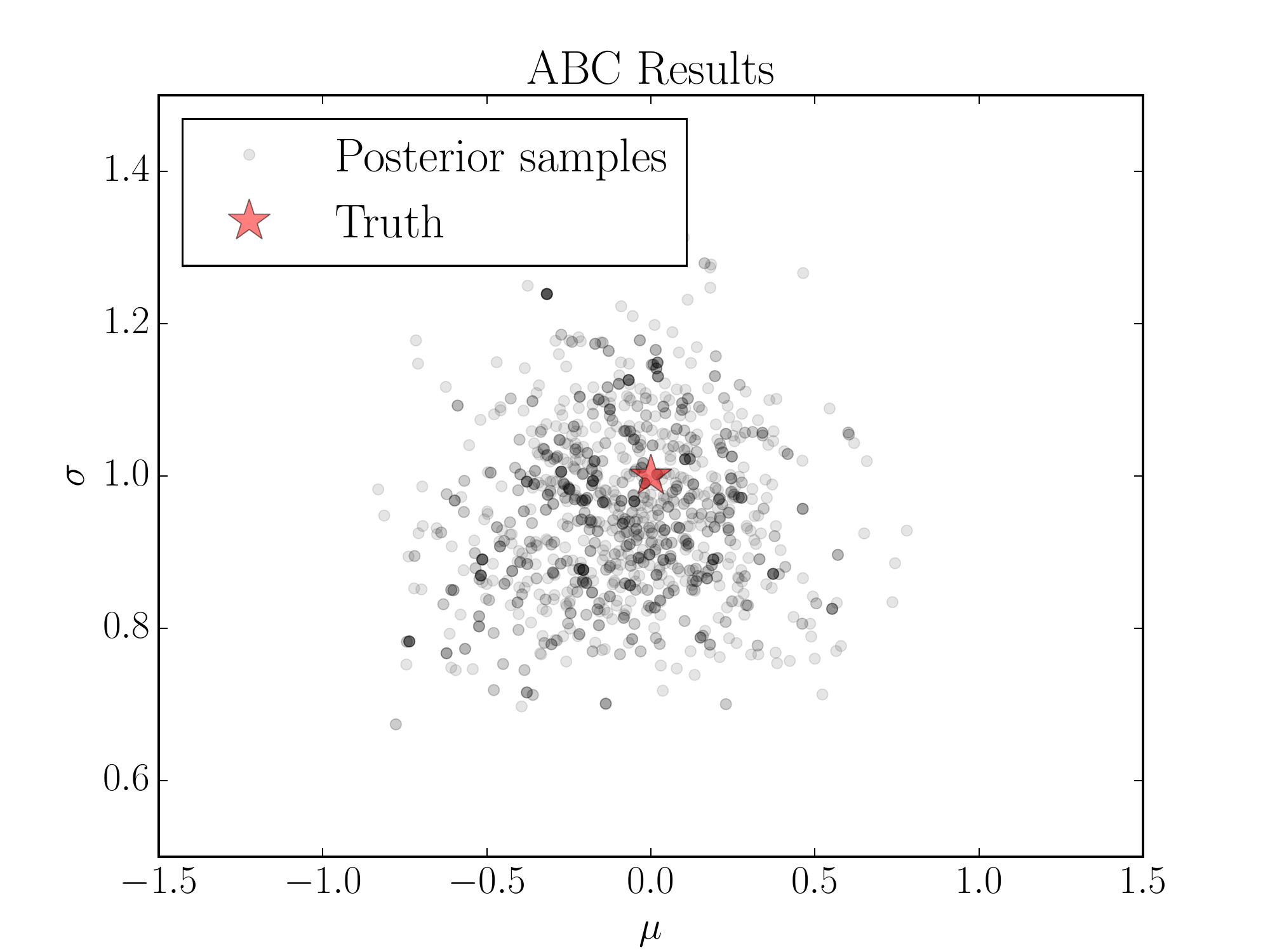}
\caption{Posterior samples for $\mu$ and $\sigma$ for the
ABC demonstration.\label{fig:abc_results}}
\end{figure}

In the example directory there is a dataset generated from
a standard normal distribution. The inference should
result in a joint posterior for $\mu$ and $\sigma$ with
significant density near $(\mu=0, \sigma=1)$.
We show posterior samples in Figure~\ref{fig:abc_results}.

\section[Python bindings]{\proglang{Python} bindings}\label{sec:python_bindings}
In \pkg{DNest4}, it is also possible to specify and run models in
\proglang{Python}. Currently, the only models supported
are ones where the parameters
are a \pkg{NumPy} array of double precision quantities.
Only a single thread is supported, although parallelization at
the level of the \proglang{Python} likelihood function should work.

A Model class implementing the straight line example is given below
and included in {\tt python/examples/straightline/straightline.py}.
The member functions of this class have similar names to those
used in C++, but work slightly differently. Most notably,
an instance of the class does not represent a point in parameter
space, but represents the problem itself. The functions take
\pkg{NumPy} arrays of parameters as input, or produce them as
output, rather than having parameters as data members as in C++.

\begin{CodeChunk}
\begin{CodeInput}
class Model(object):
    """
    Specify the model in Python.
    """
    def __init__(self):
        """
        Parameter values *are not* stored inside the class
        """
        pass

    def from_prior(self):
        """
        Unlike in C++, this must *return* a numpy array of parameters.
        """
        m = 1E3 * rng.randn()
        b = 1E3 * rng.randn()
        sigma = np.exp(-10.0 + 20.0 * rng.rand())
        return np.array([m, b, sigma])

    def perturb(self, params):
        """
        Unlike in C++, this takes a numpy array of parameters as input,
        and modifies it in-place. The return value is still logH.
        """
        logH = 0.0
        which = rng.randint(3)

        if which == 0 or which == 1:
            logH -= -0.5 * (params[which] / 1E3) **2
            params[which] += 1E3 * dnest4.randh()
            logH += -0.5 * (params[which]/1E3) **2
        else:
            log_sigma = np.log(params[2])
            log_sigma += 20 * dnest4.randh()
            # Note the difference between dnest4.wrap in Python and
            # DNest4::wrap in C++. The former *returns* the wrapped value.
            log_sigma = dnest4.wrap(log_sigma, -10.0, 10.0)
            params[2] = np.exp(log_sigma)

        return logH

    def log_likelihood(self, params):
        """
        Gaussian sampling distribution.
        """
        m, b, sigma = params
        var = sigma ** 2
        return -0.5 * data.shape[0] * np.log(2 * np.pi * var)\
                - 0.5 * np.sum((data[:,1] - (m * data[:,0] + b)) **2) / var

\end{CodeInput}
\end{CodeChunk}

A sampler can then be created and run using the following code.
The \proglang{Python} sampler does not load an {\tt OPTIONS} file.
Instead, the options are provided as function arguments
when setting up the sampler.
\begin{CodeChunk}
\begin{CodeInput}
# Create a model object and a sampler
model = Model()
sampler = dnest4.DNest4Sampler(model,
                               backend=dnest4.backends.CSVBackend(".",
                                                                  sep=" "))

# Set up the sampler. The first argument is max_num_levels
gen = sampler.sample(max_num_levels=30, num_steps=1000,\
                      new_level_interval=10000,\
                      num_per_step=10000, thread_steps=100,\
                      num_particles=5, lam=10, beta=100, seed=1234)

# Do the sampling (one iteration here = one particle save)
for i, sample in enumerate(gen):
    print("# Saved {k} particles.".format(k=(i+1)))
\end{CodeInput}
\end{CodeChunk}

The text file output can be analyzed using \code{dnest4.postprocess()}
as usual.

\section*{Acknowledgements}
John Veitch (Birmingham)
and Jorge Alarcon Ochoa (Rensselaer Polytechnic Institute)
provided helpful comments on an early
version of the manuscript.
It is a pleasure to thank Anna Pancoast (Harvard),
Ewan Cameron (Oxford), David Hogg (NYU), Daniela Huppenkothen (NYU),
and Iain Murray (Edinburgh)
for valuable discussions. This work was supported by a Marsden Fast-Start grant
from the Royal Society of New Zealand.

\bibliography{references}

\appendix
\section{Command line options}\label{sec:command_line_options}
The following command line options are available.\\

\code{-o <filename>}\\
This option loads the \pkg{DNest4} options from the specified file, allowing
an alternative to the default OPTIONS.\\

\code{-s <seed>}\\
This seeds the random number generator with the specified value. If unspecified, the system time is used.\\

\code{-d <filename>}\\
Load data from the specified file, if required.\\

\code{-c <value>}\\
Standard DNS creates levels with a volume ratio of approximately
$e\approx 2.71828$. To use a different value, such as 10, use this option.
In our experience, \code{-c 10} tends to work better than the default
value on problems with a large information $\mathcal{H}$.
The output units remain in nats.
This option is incompatible with setting the maximum number of levels to 0
(automatic) in the OPTIONS file.
\\

\code{-t <num\_threads>}\\
Run on the specified number of threads. The default is 1.
Be aware that the number of particles specified in the {\tt OPTIONS}
file is actually the number of particles {\em per thread}. Therefore,
if running on 8 or more cores, we recommend setting the number of particles
per thread to 1, so the number of particles equals the number of threads.

%\section{Specifying models in Python with dnest4.builder}

\section[Specifying models in Julia]{Specifying models in \proglang{Julia}}
Experimental support for model specification in the
\proglang{Julia} language \citep{julia}
is provided, with an example given in the
{\tt code/Templates/JuliaModel} directory.
Set an enviroment variable {\tt JULIA\_LIB\_PATH} to the directory on your
system that contains the libraries {\tt libjulia.so} and
{\tt sys.so}. Set another environment variable {\tt JULIA\_INCLUDE\_PATH}
to the directory where the \proglang{Julia} header files are located
(do not include the final {\tt julia} directory in this path).

The log likelihood is specified in {\tt julia\_model.jl}, as a function
called {\tt log\_likelihood} that takes a {\tt Vector::Float64} as an argument
and returns a {\tt Float64}. You should assume that the priors for the
parameters are independent Uniform(0, 1) distributions (sensible default
proposals are used). To use a different prior, transform the parameters inside
your {\tt log\_likelihood} function before using them.
You may add other functions to {\tt julia\_model.jl}, for example to load
a dataset to refer to in your {\tt log\_likelihood}. The example in
the repository is the ``SpikeSlab'' likelihood from
\citet{brewer2011diffusive}.

You can compile the model by invoking {\tt make}
from the {\tt JuliaModel} directory.
Only a single thread is supported when running \proglang{Julia}
models.

\end{document}